\documentclass[preprint]{elsarticle}
\usepackage{url}
\usepackage{graphicx}
\usepackage[font=normalsize]{caption}
\usepackage{subcaption}
\usepackage{amsmath}
\usepackage{amssymb}
\usepackage{amsfonts}
\usepackage{pgfplots}
\usepackage{hyperref}
\usepackage{verbatim}
\usepackage{algorithm}
\usepackage[noend]{algpseudocode}
\usepackage[export]{adjustbox}
\usepackage{doi}
\biboptions{sort&compress}
\usepackage{enumitem}
\pgfplotsset{compat = 1.16}
\usepackage{tikz}
\usetikzlibrary{calc}

\tikzstyle{expression}=[align=center,rectangle,draw=black!50,fill=black!20,
thick,inner sep=0.15cm,text width=3cm,font=\footnotesize]{}

\tikzstyle{expressionsmall}=[align=center,rectangle,draw=black!50,
fill=black!20,thick,inner sep=0.15cm,text width=2.0cm,font=\footnotesize]{}

\tikzstyle{expressionlarge}=[align=center,rectangle,draw=black!50,
fill=black!20,thick,inner sep=0.15cm,text width=7cm,font=\footnotesize]{}

\tikzstyle{empty}=[rectangle,text height=0.8cm,inner sep=0cm]{}

\tikzstyle{edge}=[thick]

\begin{document}
\begin{frontmatter}

\title{A Systematic Evaluation of Coding Strategies for Sparse Binary Images
	\tnoteref{t1}}

\tnotetext[t1]{This work has received funding from the European Research 
	Council (ERC) under the European Union’s Horizon 2020 research and 
	innovation programme(grant agreement no. 741215, ERC Advanced Grant 
	INCOVID)}

\author[add1]{Rahul Mohideen Kaja Mohideen}\corref{cor1}
\ead{rakaja@mia.uni-saarland.de}

\author[add1]{Pascal Peter}
\ead{peter@mia.uni-saarland.de}

\author[add1]{Joachim Weickert}
\ead{weickert@mia.uni-saarland.de}

\cortext[cor1]{Corresponding author}
\address[add1]{Mathematical Image Analysis Group, Faculty of Mathematics and 
	Computer Science, Campus E 1.7, Saarland University, 66041 Saarbr{\"u}cken,
	 Germany.}

\begin{abstract}

Inpainting-based compression represents images in terms of a sparse subset of 
its pixel data. Storing the carefully optimised positions of known data creates
a lossless compression problem on sparse and often scattered binary images. 
This central issue is crucial for the performance of such codecs.
Since it has only received little attention in the literature, 
we conduct the first  systematic investigation of this problem so far. 
To this end, we first review and compare a 
wide range of existing methods from image compression and general purpose 
coding in terms of their coding efficiency and runtime. Afterwards, an 
ablation study enables us to identify and isolate the most useful components 
of existing methods. With context mixing, we combine those ingredients into 
new codecs that offer either better compression ratios or a more favourable 
trade-off between speed and performance.
	
\end{abstract}

\end{frontmatter}

\section{Introduction}

Traditional lossy image compression methods such as the JPEG \cite{PM92} family
 of codecs or HEVC \cite{SOHW12} are transform-based: They compress images by 
 applying a cosine or wavelet transform and quantising the coefficients coarsely, 
 which constitutes a loss of information in the frequency domain. In recent 
 years, inpainting-based codecs \cite{GWWB08,BBBW09,DIK09,SPMEWB14,PHHW16,
 MMCB18,Pe19} have challenged this concept. They discard information directly
 in the spatial domain by representing the images through a sparse subset of 
 pixels also known as the inpainting mask or in short, the mask. For decoding, 
 they reconstruct the discarded information through inpainting. Especially for 
 high compression ratios, they can outperform  transform-based codecs. 

However, these approaches require careful and unconstrained optimisation of both
 the positions and values of the known data \cite{DDI06, BBBW09,MHWT+11,HSW13,
CRP14,KBPW18}. This creates a delicate compression problem: In general, storing 
optimised mask location data is expensive from an information theoretical perspective.
 We can argue that data optimised for quality contain more information
about the signal, which in turn would have higher entropy therefore making it  
 more expensive to store \cite{SW49}. Additionally, any deviations, 
 e.g.~by inexact, 
 cheaper positions, can also reduce the reconstruction quality during 
decompression. While having deviations in pixel values is feasible and many 
inpainting operators are robust under quantisation \cite{SPMEWB14, PHHW16, 
Pe19}, storing pixel locations is a much more sensitive task. Storing pixel 
locations is equivalent to storing a sparse binary image. Depending on the 
inpainting operator, the distribution of these points aligns with image 
structures or might be lacking obvious patterns (see Fig.\ref{fig:mask_table}). 

Most inpainting operators perform best for unconstrained, optimised data that
 are stored losslessly \cite{PHHW16}. Consequently, many optimisation methods 
 \cite{BBBW09,MHWT+11,HSW13,CRP14, KBPW18, DDI06} would benefit from lossless codecs 
 specifically aimed at optimised inpainting masks.  Even beyond inpainting
 -based compression, unconstrained mask codecs could be useful for other 
 applications such as storage of sparse image features (e.g.~SIFT \cite{Lo99}, 
 SURF \cite{BETV08}). However, there is only a small number of publications 
 \cite{DIK09,MMCB18, PHHW16} that directly address the lossless compression of
  unconstrained masks.

\subsection{Our Contributions}
To close this gap, we offer the first  systematic review and in-depth analysis 
of suitable compression techniques from different fields and propose best 
practice recommendations for lossless coding of unconstrained masks. Our contributions are 
three-fold:

We start with a systematic review of suitable coding strategies from different 
fields of compression. We cover dedicated positional data compression methods 
\cite{DIK09,MMCB18}, image compression codecs (JBIG \cite{JBIG93}, PNG 
\cite{B97}, JBIG2 \cite{HKMFR98}, DjVu \cite{BHHSBL98}, BPG lossless 
\cite{CSLL12}), and state-of-the-art general purpose encoders (PAQ, LPAQ 
\cite{Ma05}). In addition, we discuss how these approaches can be applied to 
different representations of the sparse binary images, such as run-length 
encoding and methods from data science like Coordinate List (COO) \cite{Sa03} 
and Compressed Sparse Row (CSR) \cite{Sa03}. 

Afterwards we perform an ablation study of context mixing methods. Such 
approaches combine information from multiple sources of already encoded 
information to predict the remaining file content, thus increasing compression 
efficiency. In a top-down approach, we evaluate which parts of complex 
state-of-the-art methods like LPAQ2 \cite{lpaq} are useful. Moreover, we use a 
bottom-up strategy to construct a novel context mixing method that is simple,
 yet effective. 

Finally we provide a methodical comparison of all existing and newly proposed 
methods from the previous two contributions w.r.t.~compression efficiency and 
runtime. In our experiments, we choose the Kodak image dataset \cite{Ko99}, 
which is widely used in compression studies. In order to obtain meaningful 
results, we consider different densities of known data (1\% through 10\%) and 
different point distributions. We take into account random masks, results from 
probabilistic sparsification \cite{MHWT+11} with homogeneous diffusion 
\cite{Ii59}, and densification \cite{APW17} with Shepard interpolation 
\cite{Sh68}. In total, this yields a database of 720 diverse real-world test 
images on which we test all of the methods that we consider. This setup enables us to 
propose best practice recommendations based on 
our analysis. 

Overall, our three contributions allow us to identify the coding methods that 
yield the best lossless compression performance as well as the best trade-off between 
speed and coding efficiency.

\subsection{Related Work}
\label{sec:related_work}

During the last decade, inpainting-based compression has become a diverse field
 of research with many conceptually different approaches. While most of the 
 existing works focus on the optimisation of known data and inpainting 
 operators, they have also contributed some concepts for efficient storage 
 methods. We distinguish four categories:

\emph{1) Regular grid approaches} restrict the positions of sparse known data 
to a fixed grid. Some codecs specify only a global grid size of Cartesian 
\cite{Pe19} or hexagonal \cite{HMWP13} grids, thus steering the density of the 
inpainting mask. Apart from a few exceptions such as RJIP (Regular grid codec 
with Joint Inpainting and Prediction) \cite{Pe19}, 
most regular grid approaches for inpainting-based compression do not offer 
competitive performance.

\emph{2) Subdivision approaches} additionally allow some adaptation to the 
image. Most of them use error-based refinement of the grid, splitting areas 
with high inpainting error into smaller subimages with a finer grid. They store
 these splitting decisions efficiently as a binary or quad tree. Originally, 
 Gali\'c et al.~\cite{GWWB08} and Livada et al.~\cite{LGZ12} employed a 
 triangular subdivision while later works \cite{PKW16, PHHW16} use the more 
 efficient rectangular subdivision by Schmaltz et al.~\cite{SPMEWB14}.

\emph{3) Semantic approaches} select known data directly based on image features
 and belong to the oldest inpainting-based compression techniques. They date 
 back to a 1988 paper by Carlsson \cite{Ca88}. Contemporary semantic codecs 
 are particularly popular for the compression of images with pronounced edges 
 \cite{AG94, Ba19} such as cartoons \cite{WZSG09, BHK10, MBWF11, ZD11}, depth
  maps \cite{GLG12, LSJO12, HMWP13}, or flow fields \cite{JPW20}. These methods 
  extract and store image edges and use those as known data for inpainting. 
  Since edges are connected structures, chain codes can be used for efficient
  encoding. 

\emph{4) Unconstrained mask approaches:} While there are numerous publications 
that deal with optimal selection of sparse known data for inpainting 
\cite{BBBW09,MHWT+11,HMHW+15,HSW13,CRP14, KBPW18,HPW15,BWD19,DB19,ACV11,BLP+17}, 
only few use them for actual compression. Hoffman et al.~\cite{HMWP13} 
primarily employ a regular grid as already mentioned, but they also rely on some 
unconstrained mask points to further enhance quality which are encoded with 
JBIG. Demaret et al.~\cite{DDI06} originally utilises a hybrid approach that 
combines a subdivision strategy with storage of exact positions. In a later 
publication \cite{DIK09}, Demaret et al.~replaced it by a more efficient 
context-based entropy coding method. Similar approaches where taken by Peter 
et al.~\cite{PHHW16} and Marwood et al.~\cite{MMCB18} who either apply different
 contexts or additional context mixing strategies. 
 
\begin{figure*} [t]
\centering
\begin{tabular}{ccc}
\includegraphics[width=0.3\textwidth]
{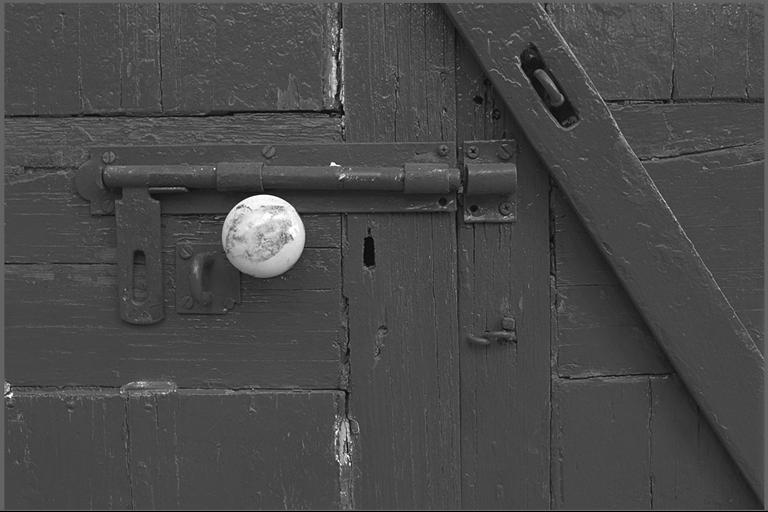} &
\includegraphics[width=0.3\textwidth]
{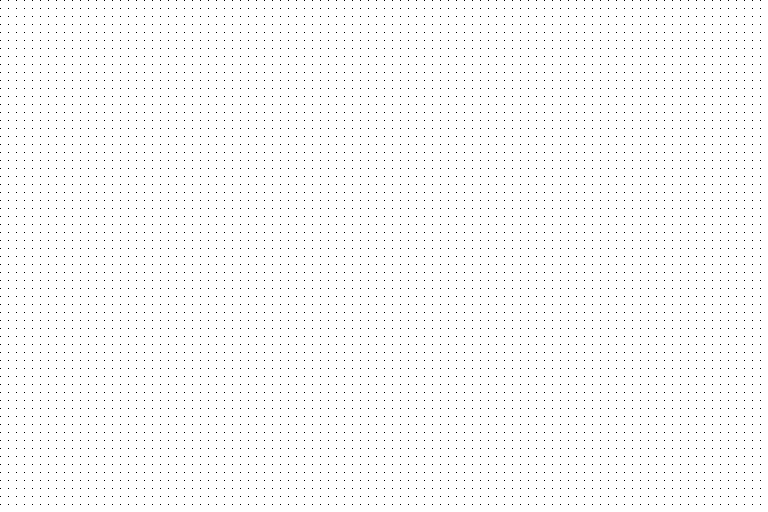} &
\includegraphics[width=0.3\textwidth]
{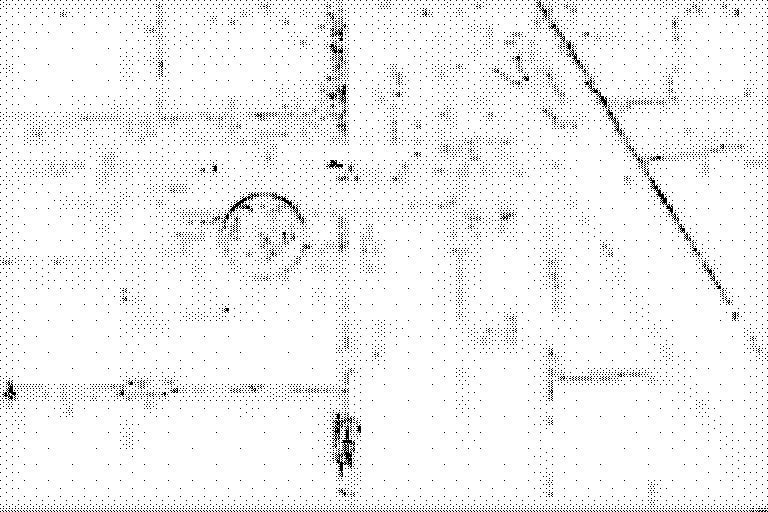} \\
original image \cite{Ko99}  & regular mask & subdivision mask \\
& & (R-EED \cite{SPMEWB14})\\
\end{tabular}
\begin{tabular}{cc}
\includegraphics[width=0.3\textwidth]
{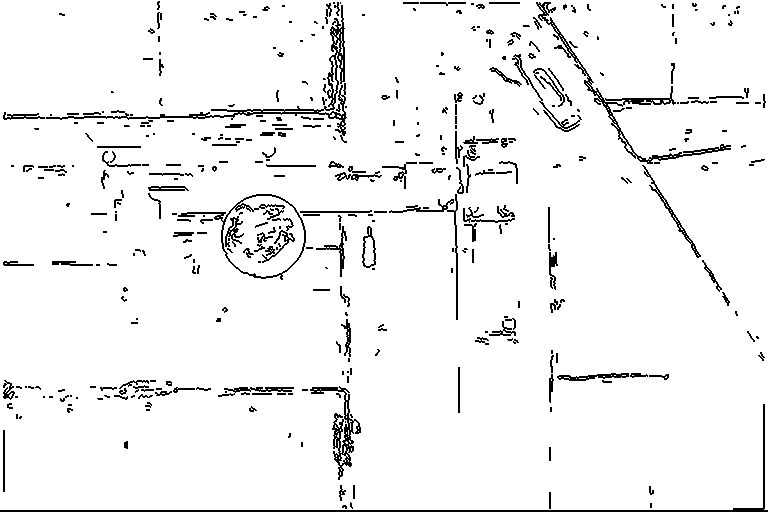} &
\includegraphics[width=0.3\textwidth]
{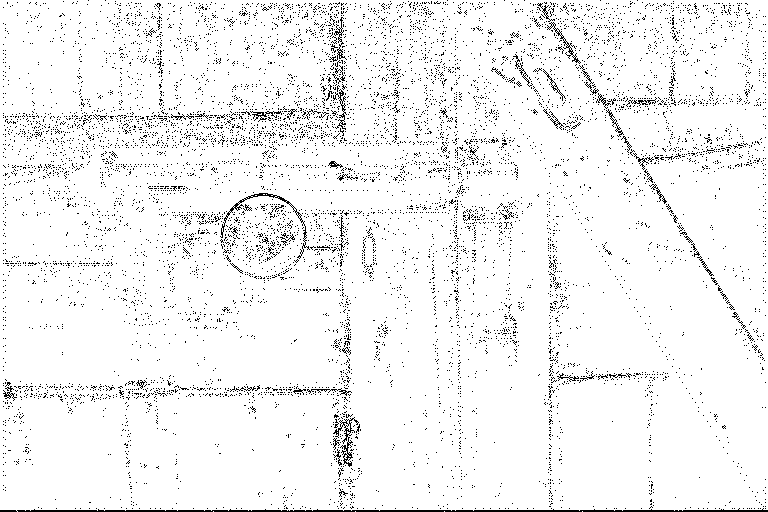} \\
edge mask (cartoon   & unconstrained mask   \\
compression \cite{MBWF11}) & (sparsification \cite{MHWT+11}) \\
\end{tabular}
\caption{In the mask images, the black pixels mark the positions of the known 
	image data. This image illustrates the mask structures we get from the 
	different approaches mentioned in Section \ref{sec:related_work}. 
	Throughout the paper, our focus will be on the compression of 
	unconstrained masks.}
\label{fig:mask_type}
\end{figure*}

 As discussed above, for almost all mask types in Figure \ref{fig:mask_type}, 
 efficient compression 
 strategies have been proposed. Unconstrained masks are the only exception.
 Therefore in the present work, 
 we restrict ourselves to the compression of unconstrained masks as they are 
 the hardest to compress out of all the different types of inpainting masks 
 that were listed.

\subsection{Structure of the Paper}

 In Section \ref{sec:inpainting_compression}, we give a short overview of 
 inpainting-based compression, which establishes the foundations for our 
 systematic analysis. We introduce several alternative approaches 
 to represent point locations in 
 Section \ref{section:rep}. We then review different entropy coding methods in Section 
 \ref{sec:encoding}. In  Section \ref{sec:image_specific}, we 
 compare different popular image compression codecs. 
 We investigate the successful ideas of context mixing in detail with 
 an ablation study in Section \ref{sec:ablation}. Furthermore, 
 to give best practice recommendations, we perform a 
 consolidated comparison of the best methods that were considered in 
 previous sections in Section \ref{section:exp}. 
 Finally, we discuss our conclusions and ideas for 
 future work in Section \ref{sec:conclusion}.

\section{Introduction to Inpainting-Based Compression}
\label{sec:inpainting_compression}

Inpainting-based compression relies on two main components: an inpainting 
method for reconstruction and a selection approach for the sparse known data 
to be stored. This can also be seen in the general structure of 
inpainting-based codecs given below in Figure \ref{fig:structure_framework}.

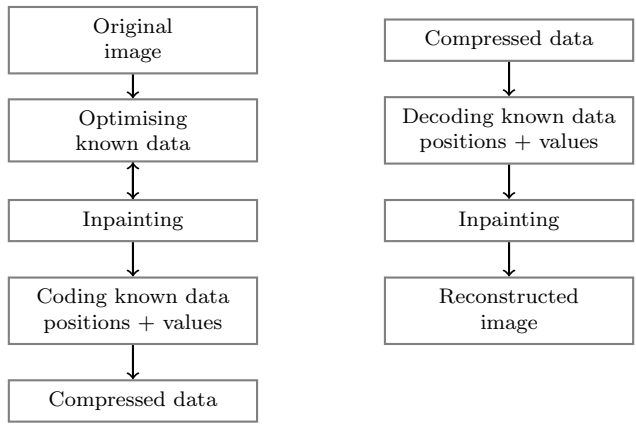
\begin{figure}[t]
	\begin{center}
		\begin{tikzpicture}
		%\node[](compression) at (-4,1.2){Compression}
		\node[expression, fill=white] (original) at (-4,0) 
		{Original \\ image};  
		\node[expression,fill=white] (selection) at (-4,-1.2)  
		{Optimising \\ known data};
		\node[expression,fill=white] (inpainting) at (-4,-2.4)  
		{Inpainting};
		\node[expression,fill=white] (coding) at (-4,-3.6)  
		{Coding known data \\ positions + values};
		\node[expression,fill=white] (compressed) at (-4,-4.8)  
		{Compressed data};
		
		\node[expression,fill=white] (compressed2) at (1,0)  
		{Compressed data};		
		\node[expression,fill=white] (decoding) at (1,-1.2) 
		{Decoding known data \\ positions + values};
		\node[expression,fill=white] (inpainting2) at (1,-2.4) 
		{Inpainting};
		\node[expression,fill=white] (recon) at (1,-3.6) 
		{Reconstructed \\ image};
		
		\draw [edge,->] (original) -- (selection);
		\draw [edge,->] (selection) -- (inpainting);
		\draw [edge,->] (inpainting) -- (selection);
		\draw [edge,->] (inpainting) -- (coding);
		\draw [edge,->] (coding) -- (compressed);
		\draw [edge,->] (compressed2) -- (decoding);
		\draw [edge,->] (decoding) -- (inpainting2);
		\draw [edge,->] (inpainting2) -- (recon);
		
		\end{tikzpicture}
		\caption{General structure of inpainting-based codecs. 
			On the compression side, we optimise the known data by repeated 
			inpaintings and adjusting the known data such that the inpainting 
			error is minimised. Then the known mask positions 
			and values are compressed using an entropy coder. The 
			compression flow of actual codecs are more complex and depend 
			on the type of mask and inpainting operator, but all of them 
			follow this general structure. On the decompression side,
			it is straightforward. We only need to decode the mask locations 
			and values and do the inpainting.  
		 }
		\label{fig:structure_framework}
	\end{center}
\end{figure}

\subsection{Inpainting Methods}
\subsubsection{Inpainting with partial differential equations}
Let $f:\Omega \rightarrow \mathbb{R}$ denote a grey value image that is only known 
on a subset $K\subset \Omega$, also called the inpainting mask. We aim to 
reconstruct the missing values in $\Omega \setminus K $ by solving for the 
steady state ($t \to \infty$) of the diffusion process \cite{We97}

\begin{align}
    \partial_t {u} &= \textnormal{div}(\boldsymbol{D}\boldsymbol{\nabla} {u}) 
    &\textnormal{on}\  \, \Omega \setminus K \times (0,\infty ),  \label{eq1}\\
    {{u}}(x,y,t)&={f}(x,y) &\textnormal{on} \ \, K\times [0,\infty),  
    \label{eq2}\\
    \boldsymbol{n}^\top \boldsymbol{D}{u}&=0 &\textnormal{on}\  \, 
    \partial\Omega \times(0,\infty).  \label{eq3} \\
     {{u}}(x,y,0)&={f}(x,y) &\textnormal{on} \ \, K\times [0,\infty),  
    \label{eq4}
\end{align}

Here, $u(x,y,t)$ is the grey value at time $t$ and position $(x,y)$, and 
$\boldsymbol{n}$ is the outer image normal. 
The spatial gradient operator is denoted by $\boldsymbol{\nabla}
=(\partial_x,\partial_y)^\top$, and  $\textnormal{div}=
 \boldsymbol{\nabla}^\top$ is the divergence. 
The diffusion tensor $\boldsymbol{D}$ is a positive definite matrix  which 
specifies the behaviour of the diffusion process. 

Eq.~\ref{eq1} models the evolution of the image over time. Eq.~\ref{eq2} and
 Eq.~\ref{eq3} define the boundary conditions for the diffusion process. The 
 Dirichlet boundary conditions in Eq.~\ref{eq2} guarantee that known pixel 
 values stay unmodified. Eq.~\ref{eq3} uses Neumann boundary conditions to 
 avoid diffusion across the image boundaries. Finally, Eq.~\ref{eq4} denotes 
 the initial condition which assigns values to the known pixels at $t=0$.

The simplest choice for the diffusion tensor is $\boldsymbol{D}=\boldsymbol{I}$
, where $\boldsymbol{I}$ is the identity matrix. In that case, Eq.~\ref{eq1} 
	can be written as
$$\partial_t {u} =\textnormal{div}(\boldsymbol{\nabla} {u}) = \Delta {u}=
\partial_{xx}{u} + \partial_{yy}{u}. $$ This equation describes homogeneous 
	diffusion as the propagation of information is equal in all directions at all 
	locations \cite{Ii59}. There are more sophisticated choices for 
$\boldsymbol{D}$ which allow e.g.~direction-dependent adaptation of the 
inpainting as in edge-enhancing anisotropic diffusion \cite{WW06}. However, a 
full review of inpainting operators is beyond the scope of our 
compression-focused contribution.

\subsubsection{Shepard interpolation}

Shepard Interpolation is a simple inpainting method that uses inverse distance 
weighting \cite{Sh68}, i.e. pixel values are inpainted by receiving more 
information from closer pixel neighbours than from those further away. 
Considering an image $f:\Omega \rightarrow \mathbb{R}$ where the inpainting 
mask is given by $K \subset \Omega$ and $\Omega$ is the image domain. We
 compute the reconstruction $u$ at a point given by the position vector 
 $\boldsymbol{p}$ as  

$$u(\boldsymbol{p}) = \frac{\sum_{\boldsymbol{q} \in K} w(\boldsymbol{p} - 
	\boldsymbol{q}) f(\boldsymbol{p})}{\sum_{\boldsymbol{q} \in K} w(
	\boldsymbol{p} - \boldsymbol{q})}.$$

We use a truncated Gaussian $G(\boldsymbol{x}):= \text{exp}(-\|\boldsymbol{x} 
\|_2^2/(2\sigma^2))$ for the weighting function $w$. The standard deviation 
$\sigma$ is adapted to the mask density according to  $\sigma = \sqrt{
	(m \cdot n)/(\pi |K|)}$ as proposed by Achanta et al.~\cite{AAS17}, where 
$m$ and $n$ are the width and height of the image respectively.

\vspace{2mm}

Beyond the two aforementioned types of inpainting methods, there are many more,
 such as exemplar-based \cite{FACS09, KBPW18} or pseudodifferential inpainting 
 \cite{AWA19}. While each of those methods has interesting properties, 
 including them into our considerations would go beyond the scope of our 
 evaluation.

\subsection{Choosing Mask Points}

While there are many different strategies to find mask points, we use two of 
the most popular ones from the literature.

\subsubsection{Probabilistic sparsification}

Sparsification \cite{MHWT+11} is a top-down algorithm where we start with the 
original image. It relies on the main idea that pixels which are harder to reconstruct are
 more important. In each iteration, the algorithm randomly chooses a set of 
 candidate pixels and removes those with the lowest per-pixel inpainting errors 
 which explains the probabilistic nature of the process. With the new mask, it
  updates the inpainting error before it proceeds to the next iteration. This 
  way, the number of mask points is reduced until the desired density is 
  achieved.
  
  \begin{figure*}[t]
  	\centering
  	\begin{tabular}{c c}
  		\includegraphics[width = 0.45\textwidth]
  		{images/original.png} & 
  		\includegraphics[width = 0.45\textwidth, frame]
  		{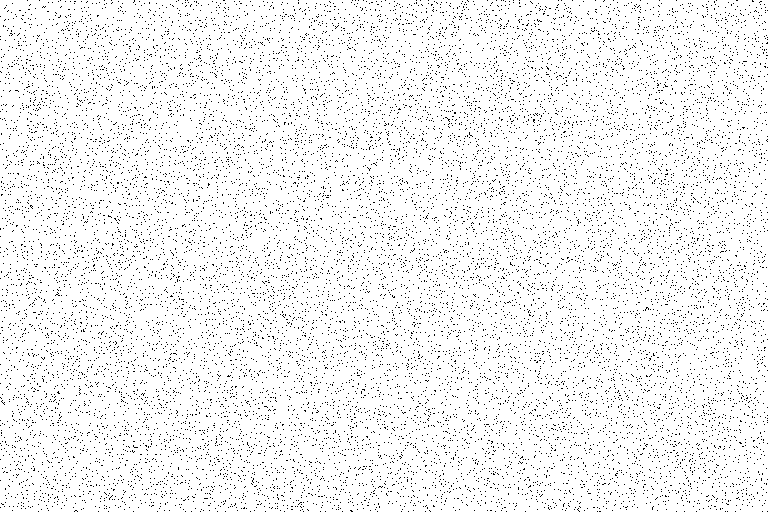} \\
  		original image \cite{Ko99} & 5\% random mask \\[2mm]
  		\includegraphics[width = 0.45\textwidth, frame]
  		{images/hom.png} & 
  		\includegraphics[width = 0.45\textwidth, frame]
  		{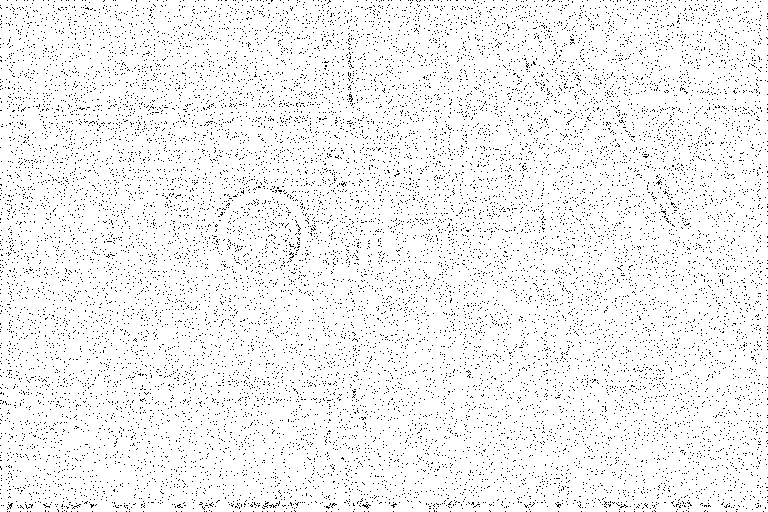} \\
  		5\% sparsification +  & 5\% densification +  \\
  		homogeneous diffusion &  Shepard interpolation \\
  	\end{tabular}
  	
  	\caption{We can see here that even though we start with the same image, 
  		different point selection strategies or inpainting operators can lead to 
  		very different unconstrained mask distributions. }
  	\label{fig:mask_table}
  \end{figure*}

\subsubsection{Probabilistic densification}
\label{susec:pd}
Densification \cite{APW17} is the bottom-up analogue of sparsification. The 
algorithm starts with an empty mask image and a fixed amount of mask pixels is 
added in every iteration until the required density is achieved. However, it has
 been found that densification generally performs worse than sparsification due
  to the fact that very little information is available in the beginning. 
For the same reason, densification tends to perform better on noisy images, 
where sparsification is prone to select noisy pixels \cite{APW17}. \par

We consider these particular methods since they generate binary masks with very
 characteristic distributions as seen in Fig.~\ref{fig:mask_table}. As a worst 
 case scenario for encoding we consider uniformly random distributions of mask 
 points. On the other side of the spectrum, we use probabilistic sparsification
  \cite{MHWT+11} and homogeneous diffusion \cite{Ii59}, which creates masks 
  that are highly adaptive to image structures. Finally, masks from 
  probabilistic densification \cite{APW17} with Shepard interpolation 
  \cite{Sh68} are a compromise between both, since they are not randomly 
  distributed, but conform less to image features. 

Unless stated otherwise, all experiments in this paper consider average 
compression performance over all the aforementioned point selection methods, 
and over all mask densities from 1\% to 10\%. 

\section{Representation of Sparse Binary images}
\label{section:rep}
In the previous section, we introduced different inpainting methods which generate the 
binary images we intend to compress. As with any other data stream, 
there are many ways to represent sparse binary
images, and the way in which the data are represented has an impact on the 
final performance of lossless compression.

Since the salient information of a sparse binary image is the non-zero pixels,
the compression ratio as a traditional metric is not the most transparent 
choice for our evaluation. To this end, we use \emph{bytes per mask pixel} 
instead, which can be computed as the compressed file size divided by the
number of non-zero pixels in the image. Such a normalisation allows us to 
attribute the cost to the salient image features that need to be stored.

In the following, we describe and 
compare common representations originating from different fields of research.

\subsection{Vector Representation}

A naive representation that allows to apply many standard sequential entropy 
coders is the vector representation. We convert the image into a vector by 
traversing it row-by-row.

\subsection{Run-length Encoding}
One of the most popular methods for compression of sparse images is run-length
encoding. It replaces sequences of identical symbols, so-called runs, by their
length.

For our setting where sparse binary images are considered, run-length encoding
becomes even more efficient. We assume that known values are mostly isolated 
and thus only have to store runs of intermediate zeroes. Consecutive ones can
be represented by a zero run.

The image has to be scanned in some order to calculate the run-lengths. 
Potential options include for example column-by-column, row-by-row,  diagonal 
or zigzag as done in JPEG. We found that the scanning order does not have a 
significant impact on the final compressed file size. Therefore, we only 
consider the most straightforward approach which is column by column. 

\begin{table}[t]
	\begin{subtable}{\textwidth}
		\centering
		\begin{tabular}{|c |c |c |c|}
			\hline
			1 & 0 & 1 & 0 \\ \hline
			0 & 0 & 0 & 1 \\ \hline
			0 & 1 & 0 & 0 \\ \hline
			0 & 0 & 1 & 0 \\   \hline
		\end{tabular}
		\\ [1ex]
		
	\end{subtable}
	
	\bigskip
	\begin{subtable}{\textwidth}
		\centering
		\begin{tabular}{|c |c |}
			\hline
			Vectorised Form & 1 0 1 0 0 0 0 1 0 1 0 0 0 0 1 0\\ \hline
			Run-length Encoding (RLE) & 0 5 1 2 1 \\
			\hline
			Coordinate List (COO) & (1,1), (1,3), (2,4), (3,2), (4,3) \\ \hline
			Compressed Sparse Row (CSR) & 1 3 4 2 3 $\mid$ 2 1 1 1 \\ \hline
			
		\end{tabular}
		\\ [1ex]
	\end{subtable}
	
	\caption{Example: binary image and its different sparse image 
		representations.}
	\label{tab:ex_rep}
\end{table}

\subsection{Coordinate List (COO)}
A COO \cite{Sa03} stores each non-zero point as a $(row, column)$ tuple. A 
differential scheme on the coordinates can also be considered, where we encode
the difference in position between the current point and the previous point. 
However, if we sort the points based on their rows and columns, we get a 
representation that is almost identical to run-length encoding.  

\subsection{Compressed Sparse Row (CSR)}
As another well known sparse image representation, CSR \cite{Sa03} is mainly 
used for fast operations on large images in scientific computing. CSR traverses
the image row-by-row and stores the column positions of each non-zero element.
In addition, the total number of non-zero elements encountered is stored when
a complete row is encoded. We make a slight modification to the basic CSR
method to yield better compression results. Instead of storing the total 
number of encountered non-zero elements, we only store the number of 
non-zero elements encountered in the previous row. This reduces the range 
of values needed to be stored thus reducing the source entropy.

%\vspace{0.5cm}
\subsection{Performance Comparisons}
\begin{figure}[t]
	\centering
	\resizebox{0.7\textwidth}{!}{
		\begin{tikzpicture}
		\begin{axis}[title style = {align=center}, xlabel = {Percentage of Mask 
			Points (\%)}, ylabel = {Bytes per Mask Pixel}, legend style = {font=\tiny}
		, ymajorgrids=true]
		\addplot[dashed,blue,mark=triangle*,mark options={fill=blue}]  
		table [x=Label, y=LPAQ, col sep=tab] {data/representation_1.csv};
		\addplot[dashed,red,mark=square*,mark options={fill=red}]  
		table [x=Label, y=LPAQ_runlength, col sep=tab] {data/representation_1.csv};
		\addplot[dashed,brown,mark=diamond*]  
		table [x=Label, y=COO, col sep=tab] {data/representation_1.csv};
		\addplot[dashed,black,mark=*,mark options={fill=black}]  
		table [x=Label, y=CSR, col sep=tab] {data/representation_1.csv};
		
		\legend{Vectorised Form, Run-length Encoding (RLE), Coordinate List (COO),
			Compressed Sparse Row (CSR)}
		\end{axis}
		\end{tikzpicture}
	}
	\caption{Compression performance of different representations compressed 
		using LPAQ2 w.r.t.~density over the mask images derived from the 
		Kodak image dataset. Vectorised form and run-length encoding are good
		choices to represent sparse matrices for compression. }
	\label{graph:rep}
\end{figure}
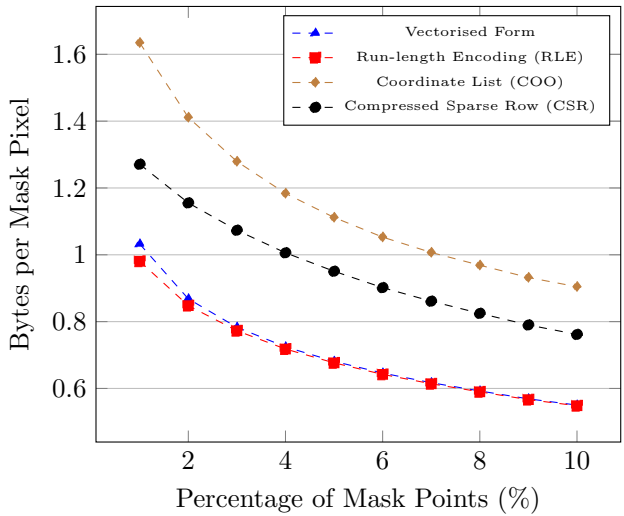

In Fig.~\ref{graph:rep}, we compare the compression performance
of different image representations w.r.t.~density by compressing them with
LPAQ. We see that the vector representation and run-length 
encoding are the better choices as they clearly lead to smaller compressed 
files than both COO and CSR.

The file size after compression depends on two aspects: the initial file size 
and the entropy of the file. The vectorised form has the same probability 
distribution as the original image. For a sparse binary image the probability 
distribution is heavily skewed towards zero, yielding a low entropy. This 
means that the original image is highly compressible which leads to high 
performance. 

Run-length encoding gives the shortest representation, as can be seen in Table
\ref{tab:ex_rep}, which in turn yields high compression performance 
especially for lower densities. On the other hand, COO and CSR have short 
representations, but not as short as RLE. Moreover, their entropy is not as low 
as the vectorised form. These factors combined make COO and CSR non-viable 
choices for compression.

\section{Introduction to Entropy Coders}
\label{sec:encoding}

In the previous section, we discussed different ways in which binary images can be represented
and their effect on the final compression performance. In this section however, we 
review approaches that constitute the final step of most image compression 
pipelines, namely entropy coders. 
Entropy coders are not specialised to image data, 
but are still a vital part and the final step of most compression codecs regardless of the
nature of the input data. The principle of 
entropy coders is to convert input data into a more compact representation which can be 
converted back to its original form by exploiting the redundancies present in 
said data. The measure of the redundancy or compressibility 
of a symbol distribution $A \rightarrow \{A_1,A_2,...,A_n\}$ with a probability 
distribution $p_1,p_2,...,p_n$ is given by the Shannon entropy \cite{SW49},
$$H(S) = \sum_i p_i \cdot log_2 (p_i).$$

Higher entropy implies that the input data has little redundancy and is harder to store.
Note that for our purposes, the input alphabet depends
on the chosen representation of the binary image from Section 
\ref{section:rep}. For a vector representation, we have $A =\{0,1\}$, and in 
the other representations $A = \{0,1,...,N\} \subset \mathbb{N}$ is a set of 
integers. 

\subsection{Huffman Coding}

Huffman coding \cite{Hu52} is a classical prefix-free encoding scheme
that maps a binary codeword to each symbol of the alphabet. As a classical 
entropy coder, Huffman coding 
  aims at distributing the code lengths in such a way that the overall coding 
  cost is minimised. 
  
 Huffman coding is simple, easy to implement, and fast. While Huffman-coding has 
 been superseded by more efficient entropy coders, it is still part of some 
 widely-used image compression codecs such as JPEG \cite{PM92} and PNG 
 \cite{B97}.
 
 \subsection{Arithmetic Coding}
 
 In contrast to Huffman coding, arithmetic coding  can encode symbols with 
 fractional bit cost per symbol, as it maps the whole source word directly to
  a binary code word.
 
 In its original form \cite{Sa18}, arithmetic coding achieves this by an 
 interval subdivision scheme.
Since this strategy requires floating-point operations and is therefore slow 
and prone to rounding errors, we consider the WNC implementation of Witten, 
Neal and Cleary \cite{WNC87}. It replaces the real-valued intervals by sets of 
integers. This greatly improves computational performance and allows an easier 
generation of the binary code word.

Due to its good approximation of the Shannon limit, arithmetic coding is used 
in many lossless and lossy image compression codecs 
\cite{SOHW12,CSLL12,JBIG93,BHHSBL98,HKMFR98,MMCB18,DIK09,GWWB08,SPMEWB14}
 and also forms the foundation for more advanced general purpose encoders such 
 as context coding and context mixing methods which we will discuss in the 
 subsequent sections.

\subsection{Context Coding}
Arithmetic coding estimates the symbol probabilities only from the symbol 
counts i.e. the number of times that the symbol has been encountered. This pure
 consideration of the occurrence frequency is referred to as a zeroth-order 
 context from which the probabilities are derived. However, more complex 
 contexts can be considered to incorporate more structural information of the 
 input data into the encoding process, thus allowing a better compression 
 performance. 

As a direct extension of the zeroth-order context, we can use a history of $m$ 
previously encoded symbols to predict the next symbol. The collection of $m$ 
previously encoded symbols is called the $m$th-order context. For images, one 
can also consider 2D neighbourhoods of varying size (see 
Fig.~\ref{context_image}). Using a single context allows us to adapt the 
probabilities to recurring patterns inside the scope of the corresponding 
contexts. This enables approaches like Prediction by Partial Matching (PPM) 
\cite{CW84}, Context Adapative Binary Arithmetic Coding (CABAC) which is used 
in HEVC \cite{SOHW12} and BPG \cite{CSLL12},
and Embedded Block Coding with Optimised Truncation 
(EBCOT) which is used in JPEG2000 \cite{CSLL12} which have better performance 
than standard arithmetic coding.

\subsection{Context Mixing}
Context mixing approaches extend upon the idea of context coding. Instead of 
using only a single context to estimate the symbol probability, 
the probabilities derived from many contexts can be combined with weighted 
averaging. 

Originating from the pioneering work of Mahoney, the so-called PAQ1 
con\-text-mixing algorithm \cite{Ma02}, many versions 
with increasingly sophisticated 
contexts have been developed. All versions of PAQ encode bits individually 
rather than the symbols themselves and convert input data into a bit stream 
accordingly. The small alphabet reduces the number of possible contexts and 
has the additional advantage that PAQ can be combined with fast binary 
arithmetic coding algorithms.

As general purpose compressors, the full versions of PAQ aim at compressing 
mixed data such as combinations of text, audio, images, and
 many other data types. Thus, many of their additional contexts would not be 
 useful for our purposes. In the following sections, we consider LPAQ2 by
  Rhatushnyak \cite{lpaq}, since we identified it as the most promising member
   of the expansive PAQ family. 

LPAQ is a lightweight variant of PAQ that uses six local contexts, and a 
prediction context. The local contexts depend on the bits already encoded in 
the current byte and the previous encoded 4 bytes. By finding the longest
repeating context, LPAQ tries to predict the next bits. A neural network mixes the 
probabilities which were obtained from the mentioned contexts. 

The neural network has a simple structure: It consists of an input layer with 
seven nodes and an output layer that computes the weighted average of the 
probabilities in the logistic domain. Such logistic mixing relies on the following 
logarithmic transformations:
\begin{align*}
  t_i &= \textnormal{ln}(p_i/(1 - p_i)) \tag*{(stretching)},\\
  p &= \sum_i w_i \cdot t_i \tag*{(mixing)},\\
    p_\text{final} &= 1/(1 + e^{-p}) \tag*{(squashing)}.
\end{align*}

The algorithm learns the weight for each context during the encoding process
 by a modified form of gradient descent that tries to minimise coding cost. 
 This results in larger weights for contexts that give good predictions for
  the given input data. The algorithm refines this probability further using 
   two SSE (Secondary Symbol Estimation) stages. Each SSE stage takes an input
  probability and then stretches and quantises them. The encoder 
  interpolates the output using a context table to give an adjusted 
  probability.  Finally, we use this probability to encode the next bit with 
  arithmetic coding. 

However, an individual context can have a different amount of influence on 
predicting the next bit depending on the local statistics. Therefore, using 
the same neural network to encode all bits can be sub-optimal. To this end, 
LPAQ2 maintains 80 neural networks and chooses which one to employ based on 
combined match lengths of the prediction context and already encoded bits.

Variants of PAQ have been utilised successfully in many inpainting-based 
compression approaches, primarily for storing quantised pixel values 
\cite{GWWB08, SPMEWB14, PKW16, HMWP13, PHHW16}. Since the PAQ family of codecs
 is so successful, we dedicate Section \ref{sec:ablation} to a detailed,
  separate analysis for them.

\section{Evaluation of Image Compression Codecs}
\label{sec:image_specific}

In Sections \ref{sec:inpainting_compression}, \ref{section:rep} and \ref{sec:encoding},
we have covered all the ingredients needed to construct sparse binary 
image compression methods. But before discussing specialised approaches to compress sparse binary images, we 
consider well-established image compression methods to provide a performance 
baseline for the specialised codecs. 

We compare five popular 
lossless image compression methods: PNG \cite{B97}, JBIG \cite{JBIG93}, 
DjVu \cite{BHHSBL98}
, JBIG2 \cite{HKMFR98}, and BPG-lossless \cite{CSLL12}.
We choose PNG as it is the most widely used lossless natural image compression method
and BPG as it is the state-of-the-art in natural image compression. We choose
JBIG and JBIG2 as they specialise in compressing binary images and finally DjVu
as it has components that are built for binary image compression.

\begin{figure}[t]
	\centering
	\resizebox{0.7\textwidth}{!}{
		\begin{tikzpicture}
		\begin{axis}[title style = {align=center}, xlabel = {Average Compression Time 
			(ms)}, ylabel = {Bytes per Mask Pixel}, legend style = {font=\small}, 
		ymajorgrids=true, legend pos= north west]
		\addplot[red, mark = o] table [x=Time, y=JBIG, col sep=tab, scatter,only 
		marks] {data/image_specific.csv};
		\addplot table [x=Time, y=PNG, col sep=tab, scatter,only marks] 
		{data/image_specific.csv};
		\addplot table [x=Time, y=BPGlossless, col sep=tab, scatter,only marks] 
		{data/image_specific.csv};
		\addplot table [x=Time, y=JBIG2, col sep=tab, scatter,only marks] 
		{data/image_specific.csv};
		\addplot table [x=Time, y=DjVu, col sep=tab, scatter,only marks] 
		{data/image_specific.csv};
		
		\legend{JBIG, PNG, BPG-lossless, JBIG2, DjVu}
		\end{axis}
		\end{tikzpicture}
	}
	\caption{Performance vs. time comparison of different image-specific 
		compression methods tested on masks derived from the Kodak image 
		dataset. PNG is the fastest and DjVu is the most efficient. JBIG and 
		JBIG2 give a good trade-off between speed and efficiency.}
	\label{graph:image_specific}
\end{figure}
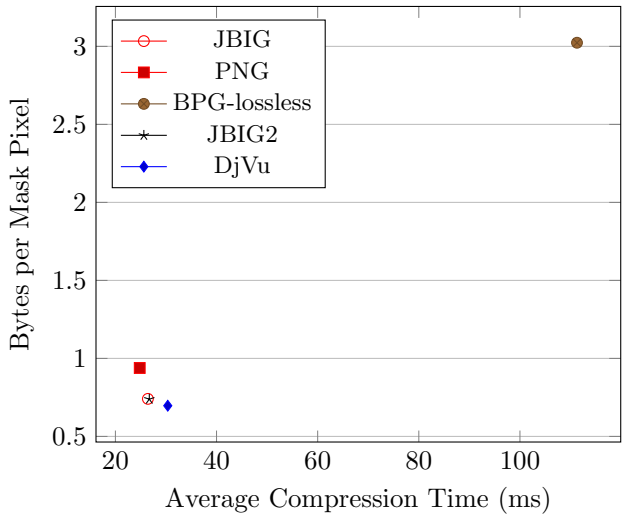

In Fig.~\ref{graph:image_specific}, we see that on masks derived from the 
Kodak dataset, PNG compresses the fastest as it combines simple prediction 
schemes with Huffman coding. DjVu performs best w.r.t.~file size reduction. 
JBIG and JBIG2 uses similar methods to compress non-text data which explains 
their comparable performance and compression times. Both give a good trade-off
between speed and performance. BPG yields inferior results because our sparse
test images violate BPG's assumptions on natural images.

\section{An Ablation Study for Context Mixing}
\label{sec:ablation}

The previous section compared several popular general-purpose image
compression methods to establish a baseline. In this section however,
we look into specialised compression methods for sparse binary images.
We have introduced 
a wide range of coding strategies that are viable candidates for good 
performance on binary images in Sections \ref{sec:encoding} 
and \ref{sec:image_specific}. However, context mixing is not only the most 
sophisticated coding strategy, it also has a track record of many applications 
in inpainting-based compression \cite{GWWB08, SPMEWB14, PKW16, HMWP13, PHHW16}. 
Therefore, we investigate which of the many components of these complex 
methods are useful for our purpose.

In the following we describe our setup for a detailed ablation study for 
context mixing on sparse binary images. In Fig.~\ref{graph:rep}, we saw that 
the vector and RLE representations were best suited for compression. Therefore, 
we pursue two approaches involving these representations: In a bottom-up 
strategy, we combine established as well as new contexts in a novel, 
minimalistic context mixing codec involving the vector representation. 
Furthermore, we deconstruct LPAQ2 in a top-down approach to compress 
run-lengths. 

\subsection{Bottom-up Context Mixing}

\label{sec:ablation_paq}

The reference point that we start with is the codec used by Marwood et 
al.~\cite{MMCB18}. It employs a global context where the probability of the next 
bit being a 1 is estimated as $V_r/N_r$ where the 
remaining number of significant pixels to be encoded is denoted by $V_r$ and 
the total remaining pixels to be coded is $N_r$. From now on, we will refer to 
this probability as the Marwood context probability. We also considered 
the context proposed by Demaret et al.~\cite{DIK09}, where the context is 
defined by the number of non-zero pixels in the 12 neighbouring pixels to the 
current pixel that were already encoded. However, we found that even though 
this performs well on its own, it does not offer any performance gain when 
combined with our models. 

\begin{figure}[t]
	\centering
	\includegraphics[width=0.3\textwidth]{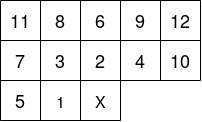}
	\caption{Local contexts for our version of PAQ. X is the pixel currently 
		being coded. The $m$th-order context corresponding to pixel X are the 
		pixels that have an index less than or equal to $m$ where $1 \leq m 
		\leq 12$.}
	
	\label{context_image}
\end{figure}

\subsubsection{BPAQ-2D-S}

We implemented a lightweight codec BPAQ-2D-S (\textbf{S}keletal \textbf{PAQ} 
for \textbf{2D} \textbf{b}inary data) inspired by the context mixing structure
  of Mahoney \cite{Ma05} where we have only the 
first order local context (neighbouring pixel to the left) in addition with 
the Marwood context. 

\begin{algorithm}
\begin{algorithmic}[1]
\item Compute evidence for the next bit being 0: 
$$S_0 = \varepsilon + w_1 n_{10}$$
\item Compute evidence for the next bit being 1:  
$$S_1 = \varepsilon + w_1 n_{11}$$
\item Compute total evidence: 
$$S = S_0 + S_1$$ 
\item Compute weighted local probability that the next bit is 1: 
$$p_{dyn} = S_1/S$$
\item Compute unweighted local probability from the first order context: 
$$p_{stat} = n_{1 1}/(n_{1 0} + n_{1 1})$$
\item Compute Marwood probability: 
$$p_{global} = V_r/N_r$$
\item Update weight for the first order context: 
$$w_1 = \max[0,w_1 + (x-p_1)(Sn_{11} - S_1 (n_{10} + n_{11}))/S_0 S_1]$$
\item Semi-stationary update of local context weight:
\begin{align*}
  n_{1x} &= n_{1x} + 1 \\
n_{1(1-x)} &= n_{1(1-x)} /2 \quad \text{if } n_{1(1-x)} > 2    
\end{align*}
\item Update counts for the Marwood probability:
\begin{align*}
V_r &= V_r - 1 \quad{\text{(if $x$ = 1)}} \\
N_r &= N_r - 1
\end{align*}
\item Compute final probability: 
$$p_{final} = 0.4 \cdot p_{dyn} + 0.2 \cdot p_{stat} + 0.4 \cdot p_{global}$$
\end{algorithmic}
\caption{BPAQ-2D-S}
\label{BPAQ-2D-S}
\end{algorithm}

The set of equations is given in Algorithm \ref{BPAQ-2D-S} where 
$n_{10}$, $n_{11}$, $w_1$ are the counts for 0 and 1 occurring in the first order
 context and its weight. The bit currently being encoded is 
denoted by $x$. A semi-stationary update for local contexts is done where we 
halve the count of the non-observed symbol. We update counts this way so that 
the context probabilities can adapt more quickly to
 local statistics \cite{Ma05}. $V_r$ is initialised to the number of non-zero 
 pixels in the image and $N_r$ is initialised to the total number of pixels in 
 the image.

When calculating $p_{\text{final}}$, we make use of a dynamically weighted local 
probability, a statically weighted local probability, and a statically 
weighted global probability. The global probability is statically weighted due 
to the fact that the counts for 0 and 1 cannot be defined for the Marwood 
context as it is done for the local context. We noticed through experimental
 observations that combining static and dynamic weights gives a better 
 performance than just using dynamic weights. We determined the static weights 
 empirically. Finally,  we encode the next bit with arithmetic coding using 
 $p_{final}$.

\subsubsection{BPAQ-2D-M}

BPAQ-2D-M (\textbf{M}ore efficient \textbf{PAQ} for \textbf{2D} 
\textbf{b}inary data) is an extended version of BPAQ-2D-S. As shown in 
Fig.~\ref{context_image}, it has twelve local contexts instead of one. 
Increasing the order from $m$ to $m+1$ adds one pixel to the 
context-neighbourhood and allows a better adaptivity to local statistics.

The difference in expressions from BPAQ-2D-S (Algorithm \ref{BPAQ-2D-S}) is 
given in Algorithm \ref{BPAQ-2D-M} where $n_{i0}$, $n_{i1}$, $w_i$ are the 
counts for 0 and 1 occurring in the $i$-th context and its corresponding 
weight, where $1 \leq i \leq 12$. The static local probability here comes from 
the fourth order context instead of the first context in BPAQ-2D-S. Again, it 
was found out experimentally that the fourth order context worked best. Apart 
from these changes, the rest of the expressions stay the same as in BPAQ-2D-S.

\begin{algorithm}
\begin{algorithmic}[1]
\item Compute evidence for the next bit being 0:
$$S_0 = \varepsilon + \sum_i w_i n_{i0}$$
\item Compute evidence for the next bit being 1:
$$S_1 = \varepsilon + \sum_i w_i n_{i1}$$
\item Compute unweighted local probability from the fourth order context: 
$$p_{stat} = n_{4 1}/(n_{4 0} + n_{4 1}) $$
\item Update context weights: 
$$w_i = \max[0,w_i + (x-p_1)(Sn_{i1} - S_1 (n_{i0} + n_{i0}))/S_0 S_1] $$
\end{algorithmic}
\caption{BPAQ-2D-M}
\label{BPAQ-2D-M}
\end{algorithm}

\subsubsection{BPAQ-2D-L}

BPAQ-2D-L (\textbf{L}ogistic \textbf{PAQ} for \textbf{2D} \textbf{b}inary data)
 is similar to BPAQ-2D-M that uses logistic mixing instead of linear mixing 
 where the probabilities are first stretched and then the final probability is 
 squashed. Unlike BPAQ-2D-M where the final probability is obtained by 
 statically mixing the mixed local context probability and the Marwood context 
 probability, here the final probability is calculated dynamically where the 
 weights change in each step for all contexts. 

In Algorithm \ref{BPAQ-2D-L}, $p_1$,\ldots,$p_{12}$ are the probabilities
 from the 12 local contexts and $p_{13}$ from the Marwood context. 
 $n_{i0}$, $n_{i1}$, $w_i$ are the counts for 0 and 1 occurring in the $i$-th 
 context and its weight. The bit currently being encoded is denoted by $x$, 
 and the learning rate $\alpha$ is set to $0.02$.

\begin{algorithm}
\begin{algorithmic}[1]
\item Compute context probabilities:
$$p_i = n_{i1}/(n_{i0} + n_{i1})$$
\item Stretch context probabilities:
$$t_i = log(p_i/(1-p_i))$$
\item Combine stretched probabilities:
$$p = \sum_i w_i \cdot t_i$$
\item Compute the final probability by squashing:
$$p = 1/(1 + e^{-p})$$
\item Update context weights:
$$w_i = w_i + \alpha \cdot t_i \cdot (x - p)$$
\item Semi-stationary update of local context counts:
\begin{align*}
    n_{ix} &= n_{ix} + 1 \\
    n_{i(1-x)} &= n_{i(1-x)} /2 \quad \text{if } n_{i(1-x)} > 2
\end{align*}
\item Update counts for the Marwood probability:
\begin{align*}
V_r &= V_r - 1 \quad{\text{(if $x$ = 1)}} \\
N_r &= N_r - 1
\end{align*}
\end{algorithmic}
\caption{BPAQ-2D-L}
\label{BPAQ-2D-L}
\end{algorithm}

\subsubsection{BPAQ-2D-XL}
Finally, we have BPAQ-2D-XL (e\textbf{X}tended \textbf{L}ogistic \textbf{PAQ} 
for \textbf{2D} \textbf{b}inary data) which is derived from BPAQ-2D-L where the 
only change is the structure of the local contexts. In the previous models, 
the higher order contexts were always contiguously extended from the lower 
context as seen in Fig.~\ref{context_image}. Here we removed the contiguity 
conditions and generated contexts that could be disconnected. We only consider 
the four nearest pixels to the current pixels, and consequentially we 
have four different first order contexts, six different second order contexts, 
four different third order contexts and one fourth order context. 
Therefore, we have 15 local contexts in total. The main differences between 
the proposed bottom-up PAQ models are listed in Table \ref{tab:ex_PAQ}.

\subsubsection{Evaluation: performance vs. runtime}
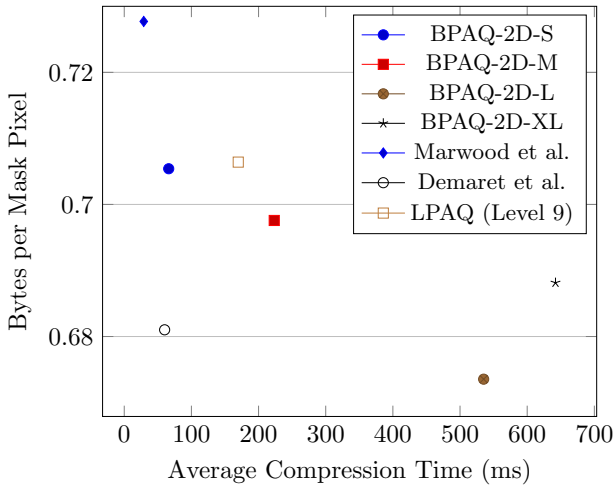
\begin{figure}[t]
    \centering
    \resizebox{0.7\textwidth}{!}{
    \begin{tikzpicture}
\begin{axis}[title style = {align=center}, xlabel = {Average Compression Time 
	(ms)}, ylabel = {Bytes per Mask Pixel}, legend style = {font=\small}, 
ymajorgrids=true, ymax = 0.73]

\addplot table [x=Time, y=PAQ_1_context, col sep=tab, scatter,only marks] 
{data/ablation_paq_1.csv};
\addplot table [x=Time, y=PAQ_12_context_static, col sep=tab, scatter,only 
marks] {data/ablation_paq_1.csv};
\addplot table [x=Time, y=PAQ_full_update_logistic, col sep=tab, scatter,
only marks] {data/ablation_paq_1.csv};
\addplot table [x=Time, y=PAQ_comb, col sep=tab, scatter,only marks] 
{data/ablation_paq_1.csv};
\addplot table [x=Time, y=Marwood, col sep=tab, scatter,only marks] 
{data/ablation_paq_1.csv};
\addplot [mark = o] table [x=Time, y=Demaret, col sep=tab, scatter,only marks] 
{data/ablation_paq_1.csv};
\addplot[brown,mark = square] table [x=Time, y=LPAQ, col sep=tab, scatter,
only marks] {data/ablation_paq_1.csv};
\legend{BPAQ-2D-S, BPAQ-2D-M, BPAQ-2D-L, BPAQ-2D-XL, Marwood et al., Demaret 
	et al., LPAQ (Level 9)}
\end{axis}
\end{tikzpicture}
}
    \caption{Performance vs.~time comparison of different PAQ based models 
    	that were implemented as part of the ablation study and the codecs 
    	proposed by Marwood et al.~\cite{MMCB18}, Demaret et al.~\cite{DIK09} 
    	and LPAQ2 applied on the image for reference. BPAQ-2D-S is the fastest 
    	and BPAQ-2D-L is the most efficient. The method of Demaret et al.~gives
    	 a good trade-off between speed and efficiency.}
    \label{graph:ablation_paq}
\end{figure}

In Fig.~\ref{graph:ablation_paq}, we see that the performance mostly correlates
 with the complexity of our proposed models. The simplest model, BPAQ-2D-S, is 
 the fastest one and BPAQ-2D-XL the slowest. Surprisingly, BPAQ-2D-L 
 outperforms all other approaches. 

BPAQ-2D-M performs better w.r.t.~compression ratio than BPAQ-2D-S due to the 
additional contexts at the cost of speed. BPAQ-2D-L gains additional 
compression efficiency, but the use of logarithmic and exponential functions 
in logistic mixing slows down the method further. The additional contexts of 
BPAQ-2D-XL do not only increase its complexity, but even slightly increase the 
cost per mask point. Thus, one should either choose BPAQ-2D-L for the lowest 
bit cost per mask point or the method of Demaret et al.~\cite{DIK09} for a 
good mix of compression ratio and speed.

\begin{table}[t]

\begin{subtable}{\textwidth}
    \centering
    \begin{tabular}{|c |c |}
    \hline
    BPAQ-2D-S & Uses 1 local context with linear mixing\\ \hline
    BPAQ-2D-M & Uses 12 local contexts with linear mixing\\ \hline
    BPAQ-2D-L & BPAQ-2D-M but with logistic mixing \\ \hline
    BPAQ-2D-XL & BPAQ-2D-L but with non-contiguous contexts \\ \hline

    \end{tabular}
    \\ [1ex]
    
    %\caption{ for the example binary image}
    \end{subtable}

    \caption{Summary of the four proposed bottom-up PAQ approaches}
    \label{tab:ex_PAQ}
\end{table}

\subsection{Top-Down Approaches}
\label{sec:ablation_lpaq}
LPAQ2 is a very good entropy coder for a wide range of data and already 
constitutes a lightweight version of PAQ. However, it is unclear if all of the 
individual steps of the method really contribute to its performance. In order 
to identify these parts, we eliminate features step-by-step in a top-down
 manner to see the performance and runtime behaviour.

We start out with the full original implementation of LPAQ2 with parameter 
settings for highest compression (compression level 9). From there on, we 
perform multiple reduction steps.

\begin{enumerate}[leftmargin=1.2cm,label=Step \arabic*:]
\item We remove the prediction context (no prediction model).
\item We reduce the amount of neural nets from 80 to a single one 
(single net model).
\item We retain only the intra-byte context and discard the rest 
(intra only model). 
\item For ULPAQ (ultra lite PAQ), we also discard the second SSE stage 
(RLE + ULPAQ). 
\end{enumerate}

  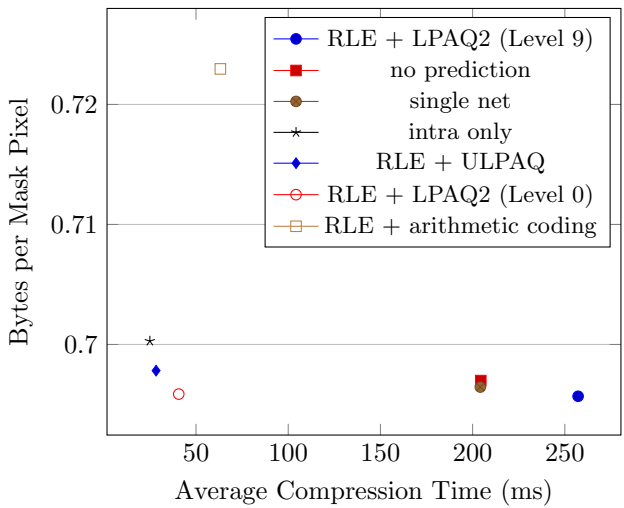
\begin{figure}[t]
    \centering
    \resizebox{0.7\textwidth}{!}{
    \begin{tikzpicture}
\begin{axis}[title style = {align=center}, xlabel = {Average Compression Time 
	(ms)}, ylabel = {Bytes per Mask Pixel}, legend style = {font=\small}, 
ymajorgrids=true, ymax = 0.728]
\addplot table [x=Time, y=Step_1, col sep=tab, scatter,only marks] 
{data/ablation_lpaq_2.csv};
\addplot table [x=Time, y=Step_2, col sep=tab, scatter,only marks] 
{data/ablation_lpaq_2.csv};
\addplot table [x=Time, y=Step_3, col sep=tab, scatter,only marks] 
{data/ablation_lpaq_2.csv};
\addplot table [x=Time, y=Step_4, col sep=tab, scatter,only marks] 
{data/ablation_lpaq_2.csv};
\addplot table [x=Time, y=Step_5, col sep=tab, scatter,only marks] 
{data/ablation_lpaq_2.csv};
\addplot[red, mark = o] table [x=Time, y=run_lpaq_0, col sep=tab, scatter,
only marks] {data/ablation_lpaq_2.csv};
\addplot [brown, mark = square]table [x=Time, y=run_arith, col sep=tab, 
scatter,only marks] {data/ablation_lpaq_2.csv};
    \legend{RLE + LPAQ2 (Level 9), no prediction, single net, intra only, 
    	RLE + ULPAQ, RLE + LPAQ2 (Level 0), RLE + arithmetic coding}
\end{axis}
\end{tikzpicture}
}
    \caption{Performance vs.~time comparison of the step-by-step removal of 
    	LPAQ2 features to compress run-lengths as part of the ablation study 
    	and LPAQ2 (Level 0) (fastest mode) and arithmetic coding for reference.
    	 RLE + ULPAQ and RLE + LPAQ2 (Level 0) gives a good trade-off between 
    	 speed and efficiency.  }
    \label{graph:ablation_lpaq}
\end{figure}

\subsubsection{Performance and runtime comparisons}

 A comparison for the aforementioned reduction steps is shown in 
 Fig.~\ref{graph:ablation_lpaq}. As the most complex model, the original LPAQ 
 is the slowest of the evaluated methods. The no prediction model and the 
 single network model only have a minor difference in bits per mask pixel, but 
 they are about 20\% faster than the original version. This shows that a single
  neural net without prediction comes at virtually no disadvantage to the more 
  complex full version of the codec.
 
 The intra only model and ULPAQ are even 10 times faster than the original 
 version. Additionally, ULPAQ performs better than intra only model. Thereby, 
 the second step of SSE is detrimental for binary input data. For reference,
  we also consider LPAQ level 0, the fastest parameter setting for standard 
  LPAQ. It is not as fast as RLE+ULPAQ, but also produces smaller encoded 
  files. In practice, ULPAQ offers a slight advantage in speed while LPAQ 
  (Level 0) provides a slight advantage in compression ratio.

\section{Best Practice Recommendations}
\label{section:exp}

In this section we perform a consolidated comparison of the best candidates for
 sparse binary mask compression from the previous Sections  \ref{section:rep}, 
 \ref{sec:encoding}, \ref{sec:image_specific} and \ref{sec:ablation}. We consider BPAQ-2D-L and the 
 codec of Demaret et al.~\cite{DIK09} from Fig.~\ref{graph:ablation_paq}, RLE 
 + ULPAQ and RLE + LPAQ2 (Level 0) from Fig.~\ref{graph:ablation_lpaq}, as well 
 as JBIG2 and DjVu from Fig.~\ref{graph:image_specific}.  

Our evaluation is based on the two essential mask characteristics: Point 
distribution and density. Both of these aspects are relevant for practical 
use. The distribution depends on the inpainting operator and mask selection 
strategy, while the density is closely connected to the desired final 
compression ratio of the inpainting method. Thus, any best practice 
recommendations need to take into account the behaviour of the algorithms
 w.r.t.~these two factors.

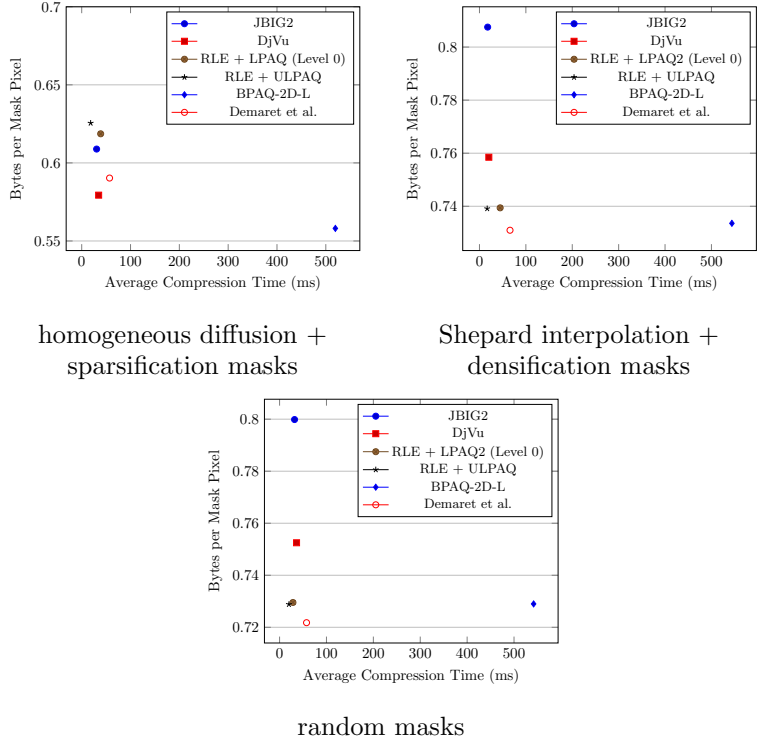
\begin{figure*}[t]
	\centering
	\begin{tabular}{c c}
\resizebox{0.4\textwidth}{!}{
		\begin{tikzpicture}
\begin{axis}[title style = {align=center}, xlabel = {Average Compression Time 
	(ms)}, ylabel = {Bytes per Mask Pixel}, legend style = {font=\small}, 
ymajorgrids=true, ymax=0.7]
\addplot table [x=Time, y=JBIG2, col sep=tab, scatter,only marks] 
{data/dist_1.csv};
\addplot table [x=Time, y=DjVu, col sep=tab, scatter,only marks] 
{data/dist_1.csv};
\addplot table [x=Time, y=LPAQ_runlength_0, col sep=tab, scatter,
only marks] {data/dist_1.csv};
\addplot table [x=Time, y=ous_LPAQ_runlength, col sep=tab, scatter,only marks] 
{data/dist_1.csv};
\addplot table [x=Time, y=Model_3, col sep=tab, scatter,only marks] 
{data/dist_1.csv};
\addplot[red, mark = o] table [x=Time, y=Demaret, col sep=tab, scatter,
only marks] {data/dist_1.csv};
\legend{JBIG2, DjVu, RLE + LPAQ (Level 0), RLE + ULPAQ, BPAQ-2D-L, 
	Demaret et al.}
\end{axis}
\end{tikzpicture}
}
& 
\resizebox{0.4\textwidth}{!}{
 \begin{tikzpicture}
\begin{axis}[title style = {align=center}, xlabel = {Average Compression Time 
	(ms)}, ylabel = {Bytes per Mask Pixel}, legend style = {font=\small}, 
ymajorgrids=true]
\addplot table [x=Time, y=JBIG2, col sep=tab, scatter,only marks] 
{data/dist_2.csv};
\addplot table [x=Time, y=DjVu, col sep=tab, scatter,only marks] 
{data/dist_2.csv};
\addplot table [x=Time, y=LPAQ_runlength_0, col sep=tab, scatter,only marks] 
{data/dist_2.csv};
\addplot table [x=Time, y=ous_LPAQ_runlength, col sep=tab, scatter,only marks] 
{data/dist_2.csv};
\addplot table [x=Time, y=Model_3, col sep=tab, scatter,only marks] 
{data/dist_2.csv};
\addplot [red, mark = o] table [x=Time, y=Demaret, col sep=tab, scatter,only 
marks] {data/dist_2.csv};
\legend{JBIG2, DjVu, RLE + LPAQ2 (Level 0), RLE + ULPAQ, BPAQ-2D-L, 
	Demaret et al.}
\end{axis}
\end{tikzpicture}
} \\[2mm]
homogeneous diffusion +  & Shepard interpolation + \\
sparsification masks & densification masks
\\[2mm]
\multicolumn{2}{c}{
\resizebox{0.4\textwidth}{!}{
 \begin{tikzpicture}
\begin{axis}[title style = {align=center}, xlabel = {Average Compression Time 
	(ms)}, ylabel = {Bytes per Mask Pixel}, legend style = {font=\small}, 
ymajorgrids=true]
\addplot table [x=Time, y=JBIG2, col sep=tab, scatter,only marks] 
{data/dist_3.csv};
\addplot table [x=Time, y=DjVu, col sep=tab, scatter,only marks] 
{data/dist_3.csv};
\addplot table [x=Time, y=LPAQ_runlength_0, col sep=tab, scatter,only marks] 
{data/dist_3.csv};
\addplot table [x=Time, y=ous_LPAQ_runlength, col sep=tab, scatter,only marks] 
{data/dist_3.csv};

\addplot table [x=Time, y=Model_3, col sep=tab, scatter,only marks] 
{data/dist_3.csv};
\addplot[red, mark = o] table [x=Time, y=Demaret, col sep=tab, scatter,only 
marks] {data/dist_3.csv};
\legend{JBIG2, DjVu, RLE + LPAQ2 (Level 0), RLE + ULPAQ, BPAQ-2D-L, 
	Demaret et al.}
\end{axis}
\end{tikzpicture}
}
} \\[2mm]
\multicolumn{2}{c}{random masks}
\end{tabular}
\caption{Performance vs. time comparison of different coders tested on masks 
	derived from different methods but averaged over all densities. For 
	homogeneous diffusion masks, DjVu gives a good performance-speed trade-off 
	where BPAQ-2D-L gives the best performance. For Shepard and random masks, 
	the method of Demaret et al.~is the best choice as it is fast and performs 
	well.}
\label{fig:dist_table}
\end{figure*}

\subsection{Comparing Compression Performance based on Point Distributions}

The different methods that we considered to generate the mask points have 
varying point distributions; see Fig.~\ref{fig:mask_table}. We want to investigate if 
compression performance changes with respect to the point distribution. Here 
we present the results averaged over all densities from 1\% to 10\% for 
different point distributions.

From Fig.~\ref{fig:dist_table} we observe that the relative rankings of 
different compression methods change over different types of distributions. 
For homogeneous diffusion masks, we see that DjVu offers both fast compression 
and a very good compression ratio. If runtime is no concern, BPAQ-2D-L provides
 the lowest cost of bits per mask point. 

For Shepard and random masks, we see that the two best methods on homogeneous 
diffusion masks, BPAQ-2D-L and DjVu, perform worse. This is due to the fact 
that masks derived from homogeneous diffusion are more structured. Both 
BPAQ-2D-L and DjVu can therefore benefit from their ability to detect such 
patterns. This structure is absent from the masks derived using Shepard 
interpolation and, in the extreme case, completely random mask distributions. 
This is also why all the codecs perform better on homogeneous diffusion masks 
than on Shepard and random masks. In such a case, the approach of Demaret et 
al.~\cite{DIK09} is a good alternative.

\subsection{Comparing Compression Performance over Image Density}

Next we evaluate if the relative performance of the selected codecs differ for 
varying densities. We present the results averaged over all point distributions
 for masks having densities of 1\%, 5\%, and 10\%. 
In addition, we also consider an average over all point distributions and all 
point densities between 1\% and 10\%.

\begin{figure*}[t]
	\centering
	\begin{tabular}{c c}
\resizebox{0.4\textwidth}{!}{
		\begin{tikzpicture}
\begin{axis}[title style = {align=center}, xlabel = {Average Compression Time 
	(ms)}, ylabel = {Bytes per Mask Pixel}, legend style = {font=\small}, 
ymajorgrids=true]
\addplot table [x=Time, y=JBIG2, col sep=tab, scatter,only marks] 
{data/density_1.csv};
\addplot table [x=Time, y=DjVu, col sep=tab, scatter,only marks] 
{data/density_1.csv};
\addplot table [x=Time, y=LPAQ_runlength_0, col sep=tab, scatter,only marks] 
{data/density_1.csv};
\addplot table [x=Time, y=ours_LPAQ_runlength, col sep=tab, scatter,only marks]
 {data/density_1.csv};
\addplot table [x=Time, y=PAQ_real, col sep=tab, scatter,only marks] 
{data/density_1.csv};
\addplot[red, mark = o] table [x=Time, y=Demaret, col sep=tab, scatter,
only marks] {data/density_1.csv};
\legend{JBIG2, DjVu, RLE + LPAQ2 (Level 0), RLE + ULPAQ, BPAQ-2D-L, 
	Demaret et al.}
\end{axis}
\end{tikzpicture}
}
& 
\resizebox{0.4\textwidth}{!}{
  \begin{tikzpicture}
\begin{axis}[title style = {align=center}, xlabel = {Average Compression Time 
	(ms)}, ylabel = {Bytes per Mask Pixel}, legend style = {font=\small}, 
ymajorgrids=true]
\addplot table [x=Time, y=JBIG2, col sep=tab, scatter,only marks] 
{data/density_5.csv};
\addplot table [x=Time, y=DjVu, col sep=tab, scatter,only marks] 
{data/density_5.csv};
\addplot table [x=Time, y=LPAQ_runlength_0, col sep=tab, scatter,only marks] 
{data/density_5.csv};
\addplot table [x=Time, y=ours_LPAQ_runlength, col sep=tab, scatter,only marks]
 {data/density_5.csv};

\addplot table [x=Time, y=PAQ_real, col sep=tab, scatter,only marks] 
{data/density_5.csv};
\addplot[red, mark = o] table [x=Time, y=Demaret, col sep=tab, scatter,only
 marks] {data/density_5.csv};
\legend{JBIG2, DjVu, RLE + LPAQ2 (Level 0), RLE + ULPAQ, BPAQ-2D-L, 
	Demaret et al.}
\end{axis}
\end{tikzpicture}
} \\[2mm]
1\% masks & 5\% masks
\\[2mm]

\resizebox{0.417\textwidth}{!}{
 \begin{tikzpicture}
\begin{axis}[title style = {align=center}, xlabel = {Average Compression Time 
	(ms)}, ylabel = {Bytes per Mask Pixel}, legend style = {font=\small}, 
ymajorgrids=true]
\addplot table [x=Time, y=JBIG2, col sep=tab, scatter,only marks] 
{data/density_10.csv};
\addplot table [x=Time, y=DjVu, col sep=tab, scatter,only marks] 
{data/density_10.csv};
\addplot table [x=Time, y=LPAQ_runlength_0, col sep=tab, scatter,only marks] 
{data/density_10.csv};
\addplot table [x=Time, y=ours_LPAQ_runlength, col sep=tab, scatter,only marks]
 {data/density_10.csv};
\addplot table [x=Time, y=PAQ_real, col sep=tab, scatter,only marks] 
{data/density_10.csv};
\addplot[red, mark = o] table [x=Time, y=Demaret, col sep=tab, scatter,
only marks] {data/density_10.csv};
\legend{JBIG2, DjVu, RLE + LPAQ2 (Level 0), RLE + ULPAQ, BPAQ-2D-L, 
	Demaret et al.}
\end{axis}
\end{tikzpicture}
}
&
\resizebox{0.4\textwidth}{!}{
    \begin{tikzpicture}
\begin{axis}[title style = {align=center}, xlabel = {Average Compression Time 
	(ms)}, ylabel = {Bytes per Mask Pixel}, legend style = {font=\small}, 
ymajorgrids=true]
\addplot table [x=Time, y=JBIG2, col sep=tab, scatter,only marks]
 {data/final.csv};
\addplot table [x=Time, y=DjVu, col sep=tab, scatter,only marks] 
{data/final.csv};
\addplot table [x=Time, y=LPAQ_runlength_0, col sep=tab, scatter,only marks] 
{data/final.csv};
\addplot table [x=Time, y=ous_LPAQ_runlength, col sep=tab, scatter,only marks] 
{data/final.csv};
\addplot table [x=Time, y=Model_3, col sep=tab, scatter,only marks] 
{data/final.csv};
\addplot[red, mark = o] table [x=Time, y=Demaret, col sep=tab, scatter,only 
marks] {data/final.csv};
\legend{JBIG2, DjVu, RLE + LPAQ2 (Level 0), RLE + ULPAQ, BPAQ-2D-L, 
	Demaret et al.}
\end{axis}
\end{tikzpicture}
}
 \\[2mm]
10\% masks & averaged over 1\% to 10\% Masks
\end{tabular}
\caption{Performance vs. time comparison of different coders tested on 
	different densities averaged over masks derived from the Kodak image 
	dataset using sparsification + homogeneous diffusion, densification + 
	Shepard interpolation, and random masks.}
\label{fig:density_table}
\end{figure*}
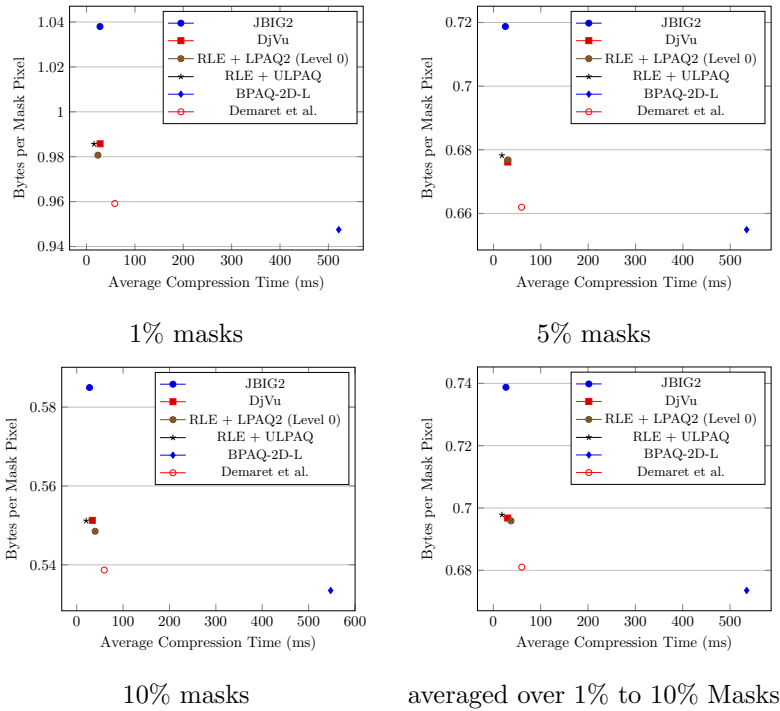

Fig.~\ref{fig:density_table} shows that the relative rankings between codecs 
are consistent over increasing densities and also in the average of all 
densities. This implies that the choice of the binary mask compression 
algorithm has to be only adapted to the distribution but not to the compression
 ratio.

For time critical applications, RLE + ULPAQ is the most suitable combination. 
BPAQ-2D-L from our ablation study of context mixing, yields the best 
compression performance on average at the cost of runtime. Between those two extremes, 
the method of Demaret et al.~provides a good balance between speed and 
performance.

Additionally, a general trend can be inferred from Fig.~\ref{graph:rep}: The 
relative cost of storing positions declines for denser masks. This shows that 
storing mask positions will decrease the compression performance by a larger 
amount for sparse masks than dense masks.

\section{Conclusions}
\label{sec:conclusion}

The first systematic study of sparse binary image compression allows us to 
understand, compare, and improve upon a wide variety of codecs. With an 
ablation study, we have determined the aspects of context mixing methods that 
make them successful for the compression of sparse binary images. As a 
consequence of the ablation study, we proposed two different classes of 
methods: one to compress the image itself and the other to compress run-lengths. 

Our best practice recommendations provide a foundation for future 
in\-pain\-ting-based codecs that use unconstrained masks. Our results from 
Section \ref{section:exp} suggest that optimal mask compression should adapt 
to the specific point distribution caused by the combination of inpainting 
operator and mask selection method. However, the compression method does not 
need to be changed based on density since that does not impact the relative 
rankings of the codecs. 

Based on our full evaluation, we have concrete recommendations for different use 
cases: RLE + ULPAQ for time-critical applications, BPAQ-2D-L for the highest amount
 of compression for structured masks, and the codec of Demaret et al.~\cite{DIK09} for 
 the compression of unstructured or random masks. The approach of Demaret et al.~can 
 also be used to compress structured masks to obtain a good 
 compromise between compression performance and speed. 

These results have interesting implications for future codecs that use 
unconstrained mask based inpainting methods. Due to the increased relative 
cost of storing very sparse unconstrained masks, we can infer that this is 
viable only when they offer a massive improvement in inpainting quality over 
regular or structured masks. It might also be interesting to consider codecs 
that produce unconstrained masks which take into account not only inpainting quality, 
but also the cost of storing them.

\bibliographystyle{elsarticle-num}
\bibliography{bib}

\begin{thebibliography}{10}
\expandafter\ifx\csname url\endcsname\relax
  \def\url#1{\texttt{#1}}\fi
\expandafter\ifx\csname urlprefix\endcsname\relax\def\urlprefix{URL }\fi
\expandafter\ifx\csname href\endcsname\relax
  \def\href#1#2{#2} \def\path#1{#1}\fi

\bibitem{PM92}
W.~B. Pennebaker, J.~L. Mitchell, {JPEG}: {S}till Image Data Compression
  Standard, Springer, New York, 1992.

\bibitem{SOHW12}
G.~J. Sullivan, J.-R. Ohm, W.J-.Han, T.~Wiegand, Overview of the high
  efficiency video coding ({HEVC}) standard, IEEE Transactions on Circuits,
  Systems and Video Technology 22~(12) (2012) 1649--1668.
\newblock \href {https://doi.org/10.1109/tcsvt.2012.2221191}
  {\path{doi:10.1109/tcsvt.2012.2221191}}.

\bibitem{GWWB08}
I.~Gali\'c, J.~Weickert, M.~Welk, A.~Bruhn, A.~Belyaev, H.-P. Seidel, Image
  compression with anisotropic diffusion, Journal of Mathematical Imaging and
  Vision 31~(2--3) (2008) 255--269.
\newblock \href {https://doi.org/10.1007/s10851-008-0087-0}
  {\path{doi:10.1007/s10851-008-0087-0}}.

\bibitem{BBBW09}
Z.~Belhachmi, D.~Bucur, B.~Burgeth, J.~Weickert, How to choose interpolation
  data in images, SIAM Journal on Applied Mathematics 70~(1) (2009) 333--352.
\newblock \href {https://doi.org/10.1137/080716396}
  {\path{doi:10.1137/080716396}}.

\bibitem{DIK09}
L.~Demaret, A.~Iske, W.~Khachabi, Contextual image compression from adaptive
  sparse data representations, in: Proc.~Workshop on Signal Processing with
  Adaptive Sparse Structured Representations, {INRIA} Rennes - Bretagne
  Atlantique, Saint Malo, France, 2009, pp. 1--6.

\bibitem{SPMEWB14}
C.~Schmaltz, P.~Peter, M.~Mainberger, F.~Ebel, J.~Weickert, A.~Bruhn,
  Understanding, optimising, and extending data compression with anisotropic
  diffusion, International Journal of Computer Vision 108~(3) (2014) 222--240.
\newblock \href {https://doi.org/10.1007/s11263-014-0702-z}
  {\path{doi:10.1007/s11263-014-0702-z}}.

\bibitem{PHHW16}
P.~Peter, S.~Hoffmann, F.~Nedwed, L.~Hoeltgen, J.~Weickert, Evaluating the true
  potential of diffusion-based inpainting in a compression context, Signal
  Processing: Image Communication 46 (2016) 40--53.
\newblock \href {https://doi.org/10.1016/j.image.2016.05.002}
  {\path{doi:10.1016/j.image.2016.05.002}}.

\bibitem{MMCB18}
D.~Marwood, P.~Massimino, M.~Covell, S.~Baluja, Representing images in 200
  bytes: Compression via triangulation, in: Proc.~25th IEEE International
  Conference on Image Processing, Athens, Greece, 2018, pp. 405--409.
\newblock \href {https://doi.org/10.1109/icip.2018.8451393}
  {\path{doi:10.1109/icip.2018.8451393}}.

\bibitem{Pe19}
P.~Peter, Fast inpainting-based compression: Combining {S}hepard interpolation
  with joint inpainting and prediction, in: Proc.~26th {IEEE} International
  Conference on Image Processing, Taipei, Taiwan, 2019, pp. 3557--3561.
\newblock \href {https://doi.org/10.1109/icip.2019.8803760}
  {\path{doi:10.1109/icip.2019.8803760}}.

\bibitem{DDI06}
L.~Demaret, N.~Dyn, A.~Iske, Image compression by linear splines over adaptive
  triangulations, Signal Processing 86~(7) (2006) 1604--1616.
\newblock \href {https://doi.org/10.1016/j.sigpro.2005.09.003}
  {\path{doi:10.1016/j.sigpro.2005.09.003}}.

\bibitem{MHWT+11}
M.~Mainberger, S.~Hoffmann, J.~Weickert, C.~H. Tang, D.~Johannsen, F.~Neumann,
  B.~Doerr, Optimising spatial and tonal data for homogeneous diffusion
  inpainting., in: A.~Bruckstein, B.~ter Haar~Romeny, A.~Bronstein,
  M.~Bronstein (Eds.), Scale Space and Variational Methods in Computer Vision,
  Vol. 6667 of Lecture Notes in Computer Science, Springer, Berlin, 2011, pp.
  26--37.
\newblock \href {https://doi.org/10.1007/978-3-642-24785-9_3}
  {\path{doi:10.1007/978-3-642-24785-9_3}}.

\bibitem{HSW13}
L.~Hoeltgen, S.~Setzer, J.~Weickert, An optimal control approach to find sparse
  data for {L}aplace interpolation, in: A.~Heyden, F.~Kahl, C.~Olsson,
  M.~Oskarsson, X.-C. Tai (Eds.), Energy Minimization Methods in Computer
  Vision and Pattern Recognition, Vol. 8081 of Lecture Notes in Computer
  Science, Springer, Berlin, 2013, pp. 151--164.
\newblock \href {https://doi.org/10.1007/978-3-642-40395-8_12}
  {\path{doi:10.1007/978-3-642-40395-8_12}}.

\bibitem{CRP14}
Y.~Chen, R.~Ranftl, T.~Pock, A bi-level view of inpainting-based image
  compression, in: Proc.~19th Computer Vision Winter Workshop, K\v{r}tiny,
  Czech Republic, 2014, pp. 19--26.

\bibitem{KBPW18}
L.~Karos, P.~Bheed, P.~Peter, J.~Weickert, Optimising data for exemplar-based
  inpainting, in: J.~Blanc-Talon, D.~Helbert, W.~Philips, D.~Popescu,
  P.~Scheunders (Eds.), Advanced Concepts for Intelligent Vision Systems,
  Lecture Notes in Computer Science, Springer, Cham, 2018, pp. 547--558.
\newblock \href {https://doi.org/10.1007/978-3-030-01449-0_46}
  {\path{doi:10.1007/978-3-030-01449-0_46}}.

\bibitem{SW49}
C.~E. Shannon, W.~Weaver, The Mathematical Theory of Communication, University
  of Illinois Press, Urbana, 1949.

\bibitem{Lo99}
D.~G. Lowe, Object recognition from local scale-invariant features, in:
  Proc.~7th IEEE International Conference on Computer Vision, Kerkyra, Greece,
  1999, pp. 1150--1157.
\newblock \href {https://doi.org/10.1109/iccv.1999.790410}
  {\path{doi:10.1109/iccv.1999.790410}}.

\bibitem{BETV08}
H.~Bay, A.~Ess, T.~Tuytelaars, L.~{Van Gool}, Speeded-up robust features
  ({SURF}), Computer Vision and Image Understanding 110~(3) (2008) 346--359.
\newblock \href {https://doi.org/10.1016/j.cviu.2007.09.014}
  {\path{doi:10.1016/j.cviu.2007.09.014}}.

\bibitem{JBIG93}
{Joint Bi-level Image Experts Group}, Information technology – progressive
  lossy/lossless coding of bi-level images, Standard, International
  Organization for Standardization (1993).

\bibitem{B97}
T.~Boutell, {PNG} ({P}ortable {N}etwork {G}raphics) specification version 1.0,
  rFC 2083, Status: Informational (Jan. 1997).
\newblock \href {https://doi.org/10.17487/rfc2083}
  {\path{doi:10.17487/rfc2083}}.

\bibitem{HKMFR98}
P.~G. Howard, F.~Kossentini, B.~Martins, S.~Forchhammer, W.~J. Rucklidge, The
  emerging {JBIG2} standard, IEEE Transactions on Circuits, Systems and Video
  Technology 8~(7) (1998) 838--848.
\newblock \href {https://doi.org/10.1109/76.735380}
  {\path{doi:10.1109/76.735380}}.

\bibitem{BHHSBL98}
L.~Bottou, P.~Haffner, P.~G. Howard, P.~Simard, Y.~Bengio, Y.~Le{C}un, High
  quality document image compression with {DjVu}, Journal of Electronic Imaging
  7~(3) (1998) 410--425.
\newblock \href {https://doi.org/10.1117/1.482609}
  {\path{doi:10.1117/1.482609}}.

\bibitem{CSLL12}
Q.~Cai, L.~Song, G.~Li, N.~Ling, Lossy and lossless intra coding performance
  evaluation: {HEVC}, {H}. 264/{AVC}, {JPEG} 2000 and {JPEG LS}, in:
  Proceedings of the 2012 Asia Pacific Signal and Information Processing
  Association Annual Summit and Conference, IEEE, 2012, pp. 1--9.

\bibitem{Ma05}
M.~Mahoney, Adaptive {W}eighing of {C}ontext {M}odels for {L}ossless {D}ata
  {C}ompression, Tech. Rep. CS-2005-16, Florida Institute of Technology,
  Melbourne, FL (Dec. 2005).

\bibitem{Sa03}
Y.~Saad, Iterative {M}ethods for {S}parse {L}inear {S}ystems, SIAM,
  Philadelphia, 2003.
\newblock \href {https://doi.org/10.1137/1.9780898718003}
  {\path{doi:10.1137/1.9780898718003}}.

\bibitem{lpaq}
M.~Mahoney, {LPAQ}, \url{http://mattmahoney.net/dc/#lpaq}, last checked:
  2020-04-08 (2007).

\bibitem{Ko99}
{Eastman Kodak Company}, \href{http://r0k.us/graphics/kodak/}{Kodak true color
  image suite}, last checked: 2020-07-24 (1999).
\newline\urlprefix\url{http://r0k.us/graphics/kodak/}

\bibitem{Ii59}
T.~Iijima, Basic theory of pattern observation, in: Papers of Technical Group
  on Automata and Automatic Control, IECE, Kyoto, Japan, 1959, pp. 3--32, in
  Japanese.

\bibitem{APW17}
R.~D. Adam, P.~Peter, J.~Weickert, Denoising by inpainting, in: F.~Lauze,
  Y.~Dong, A.~B. Dahl (Eds.), Scale Space and Variational Methods in Computer
  Vision, Vol. 10302 of Lecture Notes in Computer Science, Springer, Cham,
  2017, pp. 121--132.
\newblock \href {https://doi.org/10.1007/978-3-319-58771-4_10}
  {\path{doi:10.1007/978-3-319-58771-4_10}}.

\bibitem{Sh68}
D.~Shepard, A two-dimensional interpolation function for irregularly-spaced
  data, in: Proc.~23rd {ACM} {N}ational {C}onference, Las Vegas, NV, 1968, pp.
  517--524.
\newblock \href {https://doi.org/10.1145/800186.810616}
  {\path{doi:10.1145/800186.810616}}.

\bibitem{HMWP13}
S.~Hoffmann, M.~Mainberger, J.~Weickert, M.~Puhl, Compression of depth maps
  with segment-based homogeneous diffusion, in: A.~Kuijper, K.~Bredies,
  T.~Pock, H.~Bischof (Eds.), Scale-Space and Variational Methods in Computer
  Vision, Vol. 7893 of Lecture Notes in Computer Science, Springer, Berlin,
  2013, pp. 319--330.
\newblock \href {https://doi.org/10.1007/978-3-642-38267-3_27}
  {\path{doi:10.1007/978-3-642-38267-3_27}}.

\bibitem{LGZ12}
C.~Livada, I.~Gali\'c, B.~Zovko-Cihlar, {EEDC} image compression using
  {B}urrows-{W}heeler data modeling, in: Proc.~{Electronics in Marine}, Zadar,
  Croatia, 2012, pp. 1--5.

\bibitem{PKW16}
P.~Peter, L.~Kaufhold, J.~Weickert, Turning diffusion-based image colorization
  into efficient color compression, IEEE Transactions on Image Processing
  26~(2) (2016) 860--869.
\newblock \href {https://doi.org/10.1016/j.image.2016.05.002}
  {\path{doi:10.1016/j.image.2016.05.002}}.

\bibitem{Ca88}
S.~Carlsson, Sketch based coding of grey level images, Signal Processing 15~(1)
  (1988) 57--83.
\newblock \href {https://doi.org/10.1016/0165-1684(88)90028-x}
  {\path{doi:10.1016/0165-1684(88)90028-x}}.

\bibitem{AG94}
T.~Acar, M.~G\"okmen, Image coding using weak membrane model of images, in:
  A.~K. Katsaggelos (Ed.), Visual Communications and Image Processing, Vol.
  2308 of Proceedings of {SPIE}, SPIE Press, Bellingham, 1994, pp. 1221--1230.
\newblock \href {https://doi.org/10.1117/12.185882}
  {\path{doi:10.1117/12.185882}}.

\bibitem{Ba19}
T.~Barbu, Segmentation-based non-texture image compression framework using
  anisotropic diffusion models, Proceedings of the Romanian Academy, Series A:
  Mathematics, Physics, Technical Sciences, Information Science 20~(2) (2019)
  122--130.

\bibitem{WZSG09}
Y.~Wu, H.~Zhang, Y.~Sun, H.~Guo, Two image compression schemes based on image
  inpainting, in: Proc. 2009 International Joint Conference on Computational
  Sciences and Optimization, Sanya, China, 2009, pp. 816--820.
\newblock \href {https://doi.org/10.1109/cso.2009.470}
  {\path{doi:10.1109/cso.2009.470}}.

\bibitem{BHK10}
V.~Bastani, M.~S. Helfroush, K.~Kasiri, Image compression based on spatial
  redundancy removal and image inpainting, Journal of Zhejiang University --
  Science C (Computers \& Electronics) 11~(2) (2010) 92--100.
\newblock \href {https://doi.org/10.1631/jzus.c0910182}
  {\path{doi:10.1631/jzus.c0910182}}.

\bibitem{MBWF11}
M.~Mainberger, A.~Bruhn, J.~Weickert, S.~Forchhammer, Edge-based compression of
  cartoon-like images with homogeneous diffusion, Pattern Recognition 44~(9)
  (2011) 1859--1873.
\newblock \href {https://doi.org/10.1016/j.patcog.2010.08.004}
  {\path{doi:10.1016/j.patcog.2010.08.004}}.

\bibitem{ZD11}
C.~Zhao, M.~Du, Image compression based on {PDE}s, in: Proc.~2011 International
  Conference of Computer Science and Network Technology, Vol.~3, Harbin, China,
  2011, pp. 1768--1771.
\newblock \href {https://doi.org/10.1109/iccsnt.2011.6182311}
  {\path{doi:10.1109/iccsnt.2011.6182311}}.

\bibitem{GLG12}
J.~Gautier, O.~{Le Meur}, C.~Guillemot, Efficient depth map compression based
  on lossless edge coding and diffusion, in: Proc.~29th Picture Coding
  Symposium, Krak\'ow, Poland, 2012, pp. 81--84.
\newblock \href {https://doi.org/10.1109/pcs.2012.6213291}
  {\path{doi:10.1109/pcs.2012.6213291}}.

\bibitem{LSJO12}
Y.~Li, M.~Sjostrom, U.~Jennehag, R.~Olsson, A scalable coding approach for high
  quality depth image compression., in: Proc.~3{D}{T}{V}-Conference: {T}he True
  Vision - Capture, Transmission and Display of 3{D} Video, Zurich,
  Switzerland, 2012, pp. 1--4.
\newblock \href {https://doi.org/10.1109/3dtv.2012.6365469}
  {\path{doi:10.1109/3dtv.2012.6365469}}.

\bibitem{JPW20}
F.~Jost, P.~Peter, J.~Weickert, Compressing flow fields with edge-aware
  homogeneous diffusion inpainting, in: Proc.~45th International Conference on
  Acoustics, Speech, and Signal Processing, Barcelona, Spain, 2020, pp.
  2198--2202.
\newblock \href {https://doi.org/10.1109/icassp40776.2020.9054255}
  {\path{doi:10.1109/icassp40776.2020.9054255}}.

\bibitem{HMHW+15}
L.~Hoeltgen, M.~Mainberger, S.~Hoffmann, J.~Weickert, C.~H. Tang, S.~Setzer,
  D.~Johannsen, F.~Neumann, B.~Doerr, Optimising spatial and tonal data for
  {PDE}-based inpainting, in: M.~Bergounioux, G.~Peyr\'e, C.~Schn\"orr (Eds.),
  Variational Methods in Image Analysis, De Gruyter, Berlin, 2016, pp. 35 --
  83.
\newblock \href {https://doi.org/10.1515/9783110430394-002}
  {\path{doi:10.1515/9783110430394-002}}.

\bibitem{HPW15}
S.~Hoffmann, G.~Plonka, J.~Weickert, Discrete {Green}'s functions for harmonic
  and biharmonic inpainting with sparse atoms, in: X.-C. Tai, E.~Bae, T.~F.
  Chan, M.~Lysaker (Eds.), Energy Minimization Methods in Computer Vision and
  Pattern Recognition, Vol. 8932 of Lecture Notes in Computer Science,
  Springer, Berlin, 2015, pp. 169--182.
\newblock \href {https://doi.org/10.1007/978-3-319-14612-6_13}
  {\path{doi:10.1007/978-3-319-14612-6_13}}.

\bibitem{BWD19}
L.~Bergerhoff, J.~Weickert, Y.~Dar, Algorithms for piecewise constant signal
  approximations, in: Proc.~27th European Signal Processing Conference, IEEE,
  2019.
\newblock \href {https://doi.org/10.23919/eusipco.2019.8902559}
  {\path{doi:10.23919/eusipco.2019.8902559}}.

\bibitem{DB19}
Y.~Dar, A.~M. Bruckstein, On high-resolution adaptive sampling of deterministic
  signals, Journal of Mathematical Imaging and Vision 61~(7) (2019) 944--966.
\newblock \href {https://doi.org/10.1007/s10851-019-00880-5}
  {\path{doi:10.1007/s10851-019-00880-5}}.

\bibitem{ACV11}
G.~Albanese, M.~Cipolla, C.~Valenti, Genetic normalized convolution, in:
  G.~Maino, G.~L. Foresti (Eds.), International Conference on Image Analysis
  and Processing, Vol. 6978 of Lecture Notes in Computer Science, Springer,
  Berlin, Heidelberg, 2011, pp. 670--679.
\newblock \href {https://doi.org/10.1007/978-3-642-24085-0_68}
  {\path{doi:10.1007/978-3-642-24085-0_68}}.

\bibitem{BLP+17}
S.~Bonettini, I.~Loris, F.~Porta, M.~Prato, S.~Rebegoldi, On the convergence of
  a linesearch based proximal-gradient method for nonconvex optimization,
  Inverse Problems 33~(5) (2017) 055005.
\newblock \href {https://doi.org/10.1088/1361-6420/aa5bfd}
  {\path{doi:10.1088/1361-6420/aa5bfd}}.

\bibitem{We97}
J.~Weickert, Anisotropic Diffusion in Image Processing, Teubner, Stuttgart,
  1998.

\bibitem{WW06}
J.~Weickert, M.~Welk, Tensor field interpolation with {PDEs}, in: J.~Weickert,
  H.~Hagen (Eds.), Visualization and Processing of Tensor Fields, Springer,
  Berlin, 2006, pp. 315--325.
\newblock \href {https://doi.org/10.1007/3-540-31272-2_19}
  {\path{doi:10.1007/3-540-31272-2_19}}.

\bibitem{AAS17}
R.~Achanta, N.~Arvanitopoulos, S.~S{\"u}sstrunk, Extreme image completion, in:
  Proc.~42nd IEEE International Conference on Acoustics, Speech and Signal
  Processing, New Orleans, LA, 2017, pp. 1333--1337.
\newblock \href {https://doi.org/10.1109/icassp.2017.7952373}
  {\path{doi:10.1109/icassp.2017.7952373}}.

\bibitem{FACS09}
G.~Facciolo, P.~Arias, V.~Caselles, G.~Sapiro, Exemplar-based interpolation of
  sparsely sampled images, in: D.~Cremers, Y.~Boykov, A.~Blake, F.~Schmidt
  (Eds.), Energy Minimization Methods in Computer Vision and Pattern
  Recognition, Vol. 5681 of Lecture Notes in Computer Science, Springer,
  Berlin, 2009, pp. 331--344.
\newblock \href {https://doi.org/10.1007/978-3-642-03641-5_25}
  {\path{doi:10.1007/978-3-642-03641-5_25}}.

\bibitem{AWA19}
M.~Augustin, J.~Weickert, S.~Andris, Pseudodifferential inpainting: The missing
  link between {PDE}- and {RBF}-based interpolation, in: J.~Lellman, M.~Burger,
  J.~Modersitzki (Eds.), Scale Space and Variational Methods in Computer
  Vision, Vol. 11603 of Lecture Notes in Computer Science, Springer, Cham,
  2019, pp. 67--78.
\newblock \href {https://doi.org/10.1007/978-3-030-22368-7_6}
  {\path{doi:10.1007/978-3-030-22368-7_6}}.

\bibitem{Hu52}
D.~A. Huffman, A method for the construction of minimum redundancy codes,
  Proceedings of the IRE 40~(9) (1952) 1098--1101.
\newblock \href {https://doi.org/10.1109/jrproc.1952.273898}
  {\path{doi:10.1109/jrproc.1952.273898}}.

\bibitem{Sa18}
K.~Sayood, Arithmetic coding, in: Introduction to Data Compression, 5th
  Edition, Elsevier, 2018, Ch.~4, pp. 89--130.
\newblock \href {https://doi.org/10.1016/b978-0-12-809474-7.00004-5}
  {\path{doi:10.1016/b978-0-12-809474-7.00004-5}}.

\bibitem{WNC87}
I.~H. Witten, R.~M. Neal, J.~G. Cleary, Arithmetic coding for data compression,
  Communications of the {ACM} 30~(6) (1987) 520--540.
\newblock \href {https://doi.org/10.1145/214762.214771}
  {\path{doi:10.1145/214762.214771}}.

\bibitem{CW84}
J.~G. Cleary, I.~H. Witten, Data compression using adaptive coding and partial
  string matching, IEEE Transactions on Communications 32~(4) (1984) 396--402.
\newblock \href {https://doi.org/10.1109/tcom.1984.1096090}
  {\path{doi:10.1109/tcom.1984.1096090}}.

\bibitem{Ma02}
M.~Mahoney, The {PAQ1} {D}ata {C}ompression {P}rogram, Tech. rep., Florida
  Institute of Technology, Melbourne, FL (Mar. 2002).

\end{thebibliography}
\end{document}